\documentclass[reprint,amsmath,amssymb,aps,prc,floatfix,nofootinbib]{revtex4-2}

\usepackage[utf8]{inputenc}
\usepackage[T1]{fontenc}
\usepackage{graphicx}
\usepackage{amsmath}
\usepackage{mathtools}
\usepackage{hyperref}
\usepackage[inkscapelatex=false]{svg}
\usepackage{simplewick}
\usepackage{braket}
\DeclareMathOperator{\Tr}{Tr}
\newcommand{\refket}{|\Phi \rangle}
\newcommand{\refbra}{\langle \Phi |}
\newcommand{\expect}[1]{\big\langle #1 \big\rangle }
\newcommand{\n}{\Bar{n}}

\begin{document}

\title{In-Medium Similarity Renormalization Group at Finite Temperature}
\author{Isaac G. Smith}
\email[Corresponding author: ]{smithis7@msu.edu}
\affiliation{Facility for Rare Isotope Beams, Michigan State
University, East Lansing, MI 48824}
\affiliation{Department of Physics and Astronomy, Michigan State
University, East Lansing, MI 48824}

\author{Heiko Hergert}
\email[email: ]{hergert@frib.msu.edu}
\affiliation{Facility for Rare Isotope Beams, Michigan State
University, East Lansing, MI 48824}
\affiliation{Department of Physics and Astronomy, Michigan State
University, East Lansing, MI 48824}

\author{Scott K. Bogner}
\email[email: ]{bogner@frib.msu.edu}
\affiliation{Facility for Rare Isotope Beams, Michigan State
University, East Lansing, MI 48824}
\affiliation{Department of Physics and Astronomy, Michigan State
University, East Lansing, MI 48824}

\begin{abstract}
The study of nuclei at finite temperature is of immense interest for many areas of nuclear astrophysics and nuclear-reaction science. A variety of \textit{ab initio} methods are now available for computing the properties of nuclei from interactions rooted in Quantum Chromodynamics, but applications have largely been limited to zero temperature. In the present work, we extend one such method, the In-Medium Similarity Renormalization Group (IMSRG), to finite temperature. Using an exactly-solvable schematic model that captures essential features of nuclear interactions, we show that the FT-IMSRG can accurately determine the energetics of nuclei at finite temperature, and we explore the accuracy of the FT-IMSRG in different parameter regimes, e.g., strong and weak pairing. In anticipation of FT-IMSRG applications for finite nuclei and infinite matter, we discuss differences arising from the choice of working with the canonical and the grand canonical ensembles. In future work, we will apply the FT-IMSRG with realistic nuclear interactions to compute nuclear structure and reaction properties at finite temperature, which are important ingredients for understanding nucleosynthesis in stellar environments, or modeling reactions of hot compound nuclei.
\end{abstract}
\maketitle
\section{Introduction} \label{sec1}
Efforts to describe the properties of atomic nuclei based on nuclear forces that are rooted in Quantum Chromodynamics have made significant progress in recent decades. So-called \textit{ab initio} nuclear many-body calculations have been performed for hundreds of nuclei up to the $Z\sim 50$ region, and results for even heavier nuclei are published with increasing frequency~\cite{Hergert:2020am,Gandolfi:2020fv,Arthuis:2020la,Pastore:2020rc,Lovato:2022vy,Hu:2022sw,Lechner:2023nk,Miyagi:2024we,Door:2025fe}. 
The main challenge for most of the methods that are used in these kinds of calculations is the sheer size of the many-body Hilbert spaces, which need to encompass tens or hundreds of nucleons and their degrees of freedom, as modeled by a chosen single-particle basis. The dimension of the many-body basis scales as $\binom{N}{A}$ with the number of (indistinguishable) particles $A$ and single-particle states $N$, hence exact solutions of the (stationary) nuclear Schr\"odinger equation through diagonalization of the Hamiltonian matrix are only feasible for nuclei of mass $A\leq 20$. In order to efficiently study heavier nuclei from first principles, it is necessary to develop methods to approximate the solution to the many-body problem in polynomial time. Several such methods have been developed that can extract properties of specific energy eigenstates, most frequently the ground state (see, e.g., \cite{Hergert:2020am} and references therein).

The common first step in a large number of many-body approaches is the construction of a reference state for the many-body basis via a Hartree-Fock (HF) calculation (see, e.g., \cite{Ring:1980bb}). Hartree-Fock is a variational method to approximate the ground state of a many-body Hamiltonian as a single Slater determinant by constructing an optimal single-particle basis that minimizes the energy at the mean field level. The HF solution then serves as the basis for so-called beyond mean-field methods, which improve the approximation of the ground state in a systematic fashion and converge to the exact solution in a well-defined way. Approaches like many-body perturbation theory (MBPT) or the non-perturbative Coupled Cluster (CC) and Self-Consistent Green's Function (SCGF) methods have been very successful in approximating solutions to the nuclear many-body problem (see \cite{Hergert:2020am,Tichai:2020ft,Soma:2020lo} and references therein).

The beyond mean-field method we will focus on in this work is the In-Medium Similarity Renormalization Group (IMSRG) \cite{Tsukiyama:2011uq,Hergert:2016jk,Hergert:2017kx,Hergert:2020am}. The IMSRG applies a continuous unitary transformation to the Hamiltonian, with the goal of extracting the ground state energy, the energy of selected excited states \cite{Hergert:2017bc}, or effective interactions and operators for subsequent use in other many-body methods \cite{Stroberg:2019th,Yao:2020mw,Gebrerufael:2017fk}. Its variants have been used with great success in the prediction of ground state and excited state properties in a wide range of nuclei \cite{Hergert:2016jk,Hergert:2017kx,Stroberg:2019th,Stroberg:2021qu,Hu:2022sw,Lechner:2023nk,Miyagi:2024we,Muller:2024al}.

Thus far, applications of the IMSRG and other modern beyond-mean field methods for finite nuclei have been limited to zero temperature, although \emph{ab initio} studies of infinite matter at finite temperature based on modern nuclear interactions have been performed with a number of methods (see \cite{Drischler:2021wk} and references therein). Extensions of mean-field methods, e.g., the Finite Temperature HF (FT-HF) and Hartree-Fock-Bogoliubov (FT-HFB) \cite{Goodman:1981nx,BLaizot:1986ml,Duguet:2020ci}, and beyond mean-field methods like Shell-Model Monte Carlo \cite{Lang:1993dl,Alhassid:1994mv} have been developed decades ago, but they rely on schematic Hamiltonians or effective interactions whose parameters are fitted to data. More recently, several groups have studied nuclear ground-state properties and their response at finite temperature using both non-relativistic and relativistic energy density functionals as input \cite{Litvinova:2018df,Wibowo:2019ow,Litvinova:2020bw,Ravlic:2021wm,Ravlic:2023rd,Ravlic:2025sw}. 

In this work, we present the formalism for the finite-temperature extension of the IMSRG (FT-IMSRG), and assess its performance using an exactly solvable schematic model \cite{Hjorth-Jensen:2010qf}. Our goal is to set the stage for \emph{ab initio} calculations of nuclear properties, decay and reaction rates at finite temperatures. They are important ingredients for understanding nucleosynthesis processes in hot, stellar environments \cite{Schatz:2022ho,Goriely:2023ad}, or reactions involving hot compound nuclei, e.g., neutron-induced fission \cite{Schunck:2016fk}. Such efforts will be aided by the IMSRG's capabilities for tracking how correlations are resummed into effective in-medium interactions, akin to the spirit of nuclear Density Functional Theory (DFT) \cite{Hergert:2016jk,Duguet:2023nk}. This will allow us to link the aforementioned finite-temperature DFT work and \emph{ab initio} methodology, and provide insight into the successes and failures of either approach.

Our discussion of the FT-IMSRG is organized as follows. In Section \ref{sec2}, we present a summary of the IMSRG at zero temperature, before describing our implementation of FT-HF in Section \ref{sec3a} and the FT-IMSRG flow equations in Section \ref{sec3b}. In Section \ref{sec4}, we assess the performance of the FT-IMSRG: We introduce our schematic model in Section \ref{sec4a} and present results for a system of four fermions in eight single-particle states in Section \ref{sec4b}, before discussing larger systems in Section \ref{sec4c}. In Section \ref{sec4d}, we compare results obtained by working in the canonical and grand canonical ensembles, respectively. In Section \ref{sec4e}, we demonstrate the computation of free energy from the FT-IMSRG results. Finally, we conclude in Section \ref{sec5}. 

\section{Zero-Temperature IMSRG} \label{sec2}
\begin{figure}[t]
    \centering
    \includegraphics[width=0.48\textwidth]{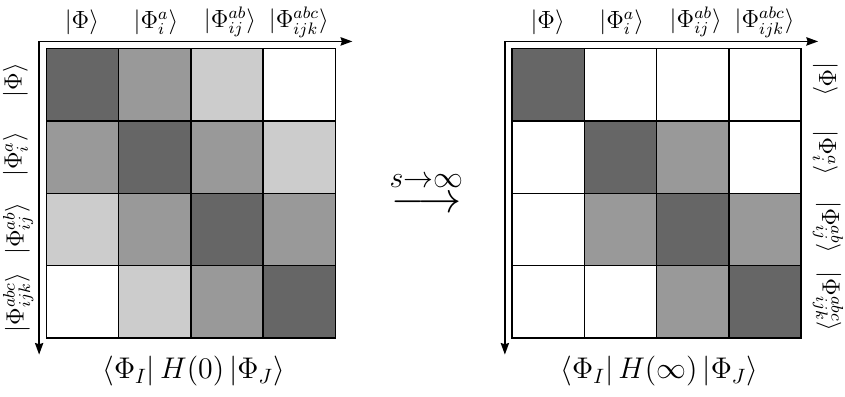}
    \caption{A schematic view of IMSRG decoupling \cite{Hergert:2016jk}. The many-body Hamiltonian is depicted in blocks, with the reference state $\refket$ and its 1-particle-1-hole excitations, 2-particle-2-hole excitations, and 3-particle-3-hole excitations (see text). During the IMSRG flow, the reference state is decoupled from its excitations.}
    \label{fig:decoupling}
\end{figure}

The first step in setting up a zero-temperature IMSRG calculation is the choice of a reference state $\refket$ \cite{Hergert:2016jk}, e.g., a HF Slater determinant serving as a first approximation to the system's ground state. A complete many-body basis can then be constructed from $\refket$, its 1-particle-1-hole excitations, 2-particle-2-hole excitations, and so on. The goal of the IMSRG is to continuously apply a unitary transformation $U(s)$ to the Hamiltonian in order to decouple $\refket$ from its excitations as $s\rightarrow \infty$, as shown schematically in Fig. \ref{fig:decoupling}. More generally, this implies that the IMSRG reshuffles correlations from the wave function into the evolved Hamiltonian so that the reference state $\refket$ becomes an eigenstate of $H(\infty)$, up to truncation errors. $U(\infty)$ can be understood as a mapping between an exact eigenstate of $H(0)$ --- usually the ground state --- and $\refket$.

To implement the IMSRG flow, we first express the Hamiltonian as
\begin{equation}\label{hf-ham}
    \hat{H} = \hat{K} + \hat{V} = \sum_{pq} k_{pq} a^\dagger_p a_q + \frac{1}{4} \sum_{pqrs} v_{pqrs} a^\dagger_p a^\dagger_q a_s a_r\,.
\end{equation}
Here, we use $\hat{K}$ for the kinetic energy to avoid confusion with the temperature later on, and we adopt chemistry conventions for labeling the HF single-particle states, where $a,b,\ldots$ refer to unoccupied (particle) states, $i,j,\ldots$ to occupied (hole) states, and $p,q,\ldots$ run over the entire basis. Next, we introduce normal-ordering with respect to the reference state $\refket$, defined by
\begin{equation}
    \{ a_p^\dagger a_q \} \equiv a_p^\dagger a_q - \refbra a_p^\dagger a_q \refket = a_p^\dagger a_q - \rho_{pq},
\end{equation}
where $\refbra a_p^\dagger a_q \refket$ is also called the (Wick) contraction of $a_p^\dagger$ and $a_q$, which is equal to the one-body density matrix $\rho_{pq}$ of the reference state. In the HF basis, this density matrix is diagonal,
\begin{equation}
    \rho_{pq}^\textrm{HF} = n_p \delta_{pq},
\end{equation}
and its eigenvalues, the occupation numbers $n_p$, are $1$ when the $p^{\rm th}$ single-particle state is occupied in the reference state (hole state) and $0$ when the $p^{\rm th}$ single-particle state is unoccupied in the reference state (particle state). A detailed description of this normal ordering scheme, including the normal ordering of general A-body operators, can be found in \cite{Hergert:2016jk,Hergert:2017kx}.

We then write the Hamiltonian in terms of normal-ordered operators:
\begin{align}
    \hat{H}&=\hat{E}+\hat{f}+\hat{\Gamma} \notag\\ &= E + \sum_{pq} f_{pq}\{a^\dagger_p a_q\} + \frac{1}{4} \sum_{pqrs} \Gamma_{pqrs}\{a^\dagger_p a^\dagger_q a_s a_r\}.
\end{align}
The 0-, 1-, and 2-body parts of the normal-ordered Hamiltonian are respectively given by
\begin{align}\label{noe}
    E &= \sum_p k_{pp}n_p + \frac{1}{2}\sum_{pq}v_{pqpq}n_pn_q \\
    \label{nof} f_{pq} &= k_{pq} + \sum_{p}v_{prqr}n_r \\
    \label{nogamma} \Gamma_{pqrs} &= v_{pqrs},
\end{align}
which can be derived directly from the n-body normal ordering in \cite{Hergert:2016jk}. We see that $E$, $f$, and $\Gamma$ now depend on the ``medium'' via the dependence on the occupation numbers of the reference state. We note that
\begin{equation}\label{e-expectation}
    E = \refbra \hat{H} \refket,
\end{equation}
making $E$ the expectation value of energy in the reference state. When we evolve the Hamiltonian as shown in Fig. \ref{fig:decoupling}, $E$ will evolve towards an exact eigenvalue of $\hat{H}$ (up to truncation effects), which is usually the desired ground-state energy. In the present work, we only deal with schematic models which include one- and two-body interactions in the initial Hamiltonian. Three-body interactions in the initial Hamiltonian can also be accounted for in the normal ordered 0-, 1-, and 2-body terms, see \cite{Hergert:2016jk,Hergert:2017kx}.

The IMSRG flow is defined through the operator differential equation
\begin{equation}\label{flow}
    \frac{d}{ds} \hat{H}(s) = [\hat{\eta}(s), \hat{H}(s)]
\end{equation}
where the generator $\hat{\eta}$ is a function of the Hamiltonian that is chosen to induce the wanted behavior out of the IMSRG flow, as will be discussed below. We note that for this flow to be unitary, the generator must be anti-Hermitian. Plugging the Hamiltonian \eqref{hf-ham} and a similarly structured $\hat{\eta}$ into the flow equation \eqref{flow} at $s=0$, we immediately find that the commutator includes three-body operators. These and further higher-rank contributions to the Hamiltonian will be induced by the IMSRG flow as we evolve. In order to close the system of flow equations, we truncate operators at the normal-ordered two-body level, in the so-called NO2B approximation. We then have the IMSRG(2) flow equations:
\begin{align}\label{flowe}
    \frac{dE}{ds} &= \sum_{pq}(n_p-n_q)\eta_{pq}f_{qp} \notag\\
    &+ \frac{1}{2}\sum_{pqrs}\eta_{pqrs}\Gamma_{rspq}n_pn_q\Bar{n}_r\Bar{n}_s \\
    \label{flowf} \frac{df_{pq}}{ds} &= \sum_r (1+P_{pq})\eta_{pr}f_{rq} \notag\\ 
    &+ \sum_{rs} (n_r-n_s)(\eta_{rs}\Gamma_{sprq}-f_{rs}\eta_{sprq}) \notag\\
    &+\frac{1}{2}\sum_{rst}(n_rn_s\Bar{n}_t + \Bar{n}_r \Bar{n}_s n_t)(1+P_{pq})\eta_{tprs}\Gamma_{rstq} \\
    \label{flowgamma}\frac{d\Gamma_{pqrs}}{ds} &= \sum_t \bigg[(1-P_{pq})(\eta_{pt}\Gamma_{tqrs}-f_{pt}\eta_{tqrs}) \notag\\ &\qquad -(1-P_{rs})(\eta_{tr}\Gamma_{pqts} -f_{tr}\eta_{pqts})\bigg] \notag \\
    &+ \sum_{tu} \bigg[\frac{1}{2}(1-n_t-n_u)(\eta_{pqtu}\Gamma_{turs}-\Gamma_{pqtu}\eta_{turs}) \notag \\
    & \qquad + (n_t-n_u)(1-P_{pq})(1-P_{rs})\eta_{tpur}\Gamma_{uqts}\bigg].
\end{align}
Here, $\Bar{n}_p = 1-n_p$ is the hole occupation number, and $P_{pq}$ is an operator which switches the $p$ and $q$ indices in subsequent expressions. Naively, with a single-particle basis size $N$, these equations scale computationally as $\mathcal{O}(N^6)$, but by differentiating between particle and hole states, it is possible to reduce this to $\mathcal{O}(N_p^4N_h^2)$, were $N_p$ and $N_h$ are respectively the number of particle and hole states~\cite{Hergert:2017bc}.

There are several generators that can be used to achieve the desired decoupling. We seek to decouple the reference state and its particle-hole excitations, which means we want to  suppress the so-called ``off-diagonal'' matrix elements of the form
\begin{equation}
    \refbra \hat{H} \{a^\dagger_a a_i\} \refket = f_{ai}
    \label{eq:f_od}
\end{equation}
and
\begin{equation}
    \refbra \hat{H} \{a^\dagger_a a^\dagger_{b} a_{j} a_i\}\refket = \Gamma_{abij} \,.\label{eq:Gamma_od}
\end{equation}
To this end, we use a generator of the form \cite{Hergert:2016jk,Hergert:2017kx}
\begin{align}
    \hat{\eta} &= \hat{\eta}^\textrm{1B} + \hat{\eta}^\textrm{2B} \notag\\ &= \sum_{ai} \eta_{ai}\{a^\dagger_a a_i\} + \frac{1}{4} \sum_{abij} \eta_{abij}\{a^\dagger_a a^\dagger_{b} a_{j} a_i\} \notag\\ &\qquad- \textrm{H.c.},
\end{align}
where $\eta_{ai}$ and $\eta_{abij}$ are related respectively to the off-diagonal matrix elements we seek to eliminate, $f_{ai}$ and $\Gamma_{abij}$.

A simple yet effective generator is the so-called White generator \cite{Hergert:2016jk,White:2002fk}
\begin{align}
    \hat{\eta}^\textrm{W} &= \sum_{ai} \frac{f_{ai}}{\Delta_{ai}}\{a^\dagger_a a_i\} + \frac{1}{4} \sum_{abij} \frac{\Gamma_{abij}}{\Delta_{abij}}\{a^\dagger_a a^\dagger_{b} a_{j} a_i\} \notag\\ &\qquad-\textrm{H.c.},
\end{align}
where
\begin{align}
    \Delta_{ai} &= \refbra \{a^\dagger_i a_a\}\hat{H}\{a^\dagger_a a_i\}\refket - \refbra \hat{H} \refket \notag\\
    &= f_{aa} - f_{ii} - \Gamma_{aiai}
\end{align}
is the unperturbed energy difference between the reference state and its p-h excitation, and
\begin{align}
    \Delta_{abij} &= \refbra \{a^\dagger_i a^\dagger_{j} a_{b} a_a\}\hat{H}\{a^\dagger_a a^\dagger_{b} a_{j} a_i\}\refket - \refbra \hat{H} \refket \notag\\
    &= f_{aa} + f_{bb} - f_{ii} - f_{jj} + \Gamma_{abab} + \Gamma_{ijij} \notag\\ &\qquad- \Gamma_{aiai} - \Gamma_{bjbj} - \Gamma_{ajaj} - \Gamma_{bibi}
\end{align}
is the unperturbed energy difference between the reference state and its 2p-2h excitation. A variation, the White arctan generator,
\begin{align}
    \hat{\eta}^\textrm{arctan} &= \frac{1}{2} \sum_{ai} \arctan\left(\frac{2f_{ai}}{\Delta_{ai}}\right) \{a^\dagger_a a_i\} \notag\\ &+ \frac{1}{8} \sum_{abij} \arctan\left(\frac{2\Gamma_{abij}}{\Delta_{abij}}\right)\{a^\dagger_a a^\dagger_{b} a_{j} a_i\} \notag\\ &\qquad- \textrm{H.c.}
\end{align}
is useful in cases where the energy denominators $\Delta_{ai}$ and $\Delta_{abij}$ become small \cite{Hergert:2017kx,White:2002fk}. Using either of these generators keeps the overall computational scaling of the IMSRG at $\mathcal{O}(N_p^4N_h^2)$.

Finally, we note that the IMSRG can determine observable quantities besides energy. Given any operator $\hat{O}$ written in the HF basis
\begin{align}
    \hat{O} &= \hat{O}^\textrm{1B} +\hat{O}^\textrm{2B} \notag \\ &= \sum_{pq} O^\textrm{1B}_{pq}a^\dagger_p a_q + \frac{1}{4} \sum_{pqrs} O^\textrm{2B}_{pqrs} a^\dagger_p a^\dagger_q a_s a_r,
\end{align}
the IMSRG flow equation for $\hat{O}$ is
\begin{equation}\label{flowq}
    \frac{d}{ds} \hat{O}(s) = [\hat{\eta}(s), \hat{O}(s)].
\end{equation}
The normal ordering (Eqs. \eqref{noe}--\eqref{nogamma}) remains unchanged, and we simply need to replace $\hat{K}$ and $\hat{V}$ with $\hat{O}^\textrm{1B}$ and $\hat{O}^\textrm{2B}$ respectively. The normal-ordered flow equations (Eqs. \eqref{flowe}--\eqref{flowgamma}) also remain unchanged, replacing $E$, $\hat{f}$ and $\hat{\Gamma}$ with the normal-ordered zero-, one- and two-body parts of $\hat{O}$. Note that the generator is the same as in the Hamiltonian flow, calculated based on the Hamiltonian's normal-ordered matrix elements, hence Eqs. \eqref{flow} and \eqref{flowq} must be solved simultaneously. Alternatively, one can use the so-called Magnus operator formulation of the IMSRG to extract a parameterization of the unitary transformation, which can then be used to construct any $\hat{O}(s)$ at a later time \cite{Morris:2015ve,Hergert:2016jk}.

\section{Finite-Temperature IMSRG} \label{sec3}
\subsection{Finite-Temperature Hartree-Fock} 
\label{sec3a}
In order to extend the zero-temperature IMSRG framework to finite temperature, we first consider the selection of a reference \emph{ensemble} for setting up our normal ordered operators. Here, we employ the finite-temperature HF (FT-HF) approach \cite{Bertsch:2016jq,Duguet:2020ci,Ryssens:2021nq}. Conventionally,  FT-HF is implemented in the  grand-canonical ensemble, whose density operator is defined as
\begin{equation}
    \hat{\rho} \equiv \frac{1}{Z}e^{-\beta(\hat{H} - \mu \hat{A})}\label{eq:gce_def_rho}
\end{equation}
with the partition function 
\begin{equation}
    Z \equiv \Tr e^{-\beta(\hat{H}-\mu \hat{A})}\,.\label{eq:gce_def_z}
\end{equation}
Here, $\hat{A}$ is the particle-number operator, $\beta\equiv1/T$ is the inverse temperature (with $k_B=1$), and $\mu$ is the chemical potential. In FT-HF, the grand partition function is approximated as
\begin{align}
  Z \approx
      \sum_{\{m_i\}} \bra{\{m_i\}} e^{-\beta\sum_p(\epsilon_p - \mu)a^\dagger_pa_p} \ket{\{m_i\}}\,,\label{eq:gce_z}
\end{align}
where $\epsilon_p$ are the single-particle energies associated with the single-particle basis that spans the occupation number states $\ket{\{m_i\}}$. For a fermion system, the $\ket{\{m_i\}}$ are Slater determinants with $m_i\in\{0,1\}$. Since they are eigenstates of the exponentiated one-body operator in Eq.~\eqref{eq:gce_z}, the approximate partition function factorizes, and we can carry out the sum over the occupation numbers to obtain
\begin{equation}
   Z^\text{HF} \sim \prod_{p}\left(1 + e^{-\beta(\epsilon_p - \mu)}\right)\,, 
\end{equation}
as well as the usual Fermi-Dirac distribution for the thermal average occupation numbers:
\begin{equation}\label{eq:gce_occ}
    n_p = \Tr(\hat{\rho} a^\dag_p a_p) = \frac{1}{1+e^{\beta (e_p-\mu)}}\,,\quad \bar{n}_p \equiv 1-n_p\,.
\end{equation}
The optimal single-particle energies $\epsilon_p$ and associated single-particle wave functions are determined by minimizing the grand potential
\begin{equation}
 \Omega = E - TS -\mu A\,,
\end{equation}
adjusting the chemical potential so that the constraint
\begin{equation}\label{eq:gce_constraint}
    \sum_p n_p = A
\end{equation}
is satisfied. The resulting optimal single-particle energies are given by
\begin{equation}
    \epsilon_p = k_{pp} + \sum_{q}v_{pqpq}n_q\,,\label{eq:gce_fthf_eps}
\end{equation}
the (FT-)HF internal energy is given by
\begin{equation}
    E = \sum_p k_p n_p + \frac{1}{2}\sum_{pq}v_{pqpq} n_p n_q \equiv \sum_p \epsilon_p n_p - V_\text{HF} \label{eq:gce_fthf_energy}
\end{equation}
and the entropy is
\begin{equation}
    S^\text{HF} = - \sum_p \left(n_p \ln n_p + \overline{n}_p \ln \overline{n}_p\right)\,.
    \label{eq:gce_fthf_entropy}
\end{equation}
Since the $\epsilon_p$ double count the contributions from the two-body interaction, Eq.~\eqref{eq:gce_fthf_energy} is used to obtain the final form of the partition function \cite{Fanto:2017ei,Ryssens:2021nq},
\begin{equation}
   Z^\text{HF} = e^{\beta V_\text{HF}}\prod_{p}\left(1 + e^{-\beta(\epsilon_p - \mu)}\right)\,. \label{eq:gce_fthf_partition}
\end{equation}

The grand-canonical FT-HF method is computationally efficient due to the factorized form of $Z^\text{HF}$: For any fixed temperature $T$, the cost matches that of a constrained zero-temperature HF calculation, which scales as $\mathcal{O}(N^4)$ in the single-particle basis size $N$. From a physical perspective, however, the grand-canonical ensemble only provides an imperfect description of atomic nuclei, since the system can exchange particles with its environment, and only the average particle number, i.e., the expectation value $\expect{\hat{A}}$, is fixed by the constraint \eqref{eq:gce_constraint}. The more appropriate choice would be the canonical ensemble, defined by
\begin{equation}
    \hat{\rho}_A \equiv \frac{1}{Z_A}e^{-\beta\hat{H}}\,, \quad Z_A \equiv \Tr e^{-\beta\hat{H}}\,,\label{eq:ce_def_rho}
\end{equation}
which describes a system with fixed particle number that can exchange energy with an environment that is at a temperature $T$. The canonical partition function is approximated as
\begin{equation} 
  Z_A \approx
      \sum^{m_1+\ldots m_N = A}_{\{m_i\}} \bra{\{m_i\}} e^{-\beta\epsilon_p\,a^\dagger_pa_p} \ket{\{m_i\}}\,.
\end{equation}
In contrast to the grand-canonical case (cf.~Eq.~\eqref{eq:gce_z}, the particle-number condition prevents us from summing over occupation numbers independently, and forces us to explicitly perform the trace over a basis of $\binom{N}{A}$ Slater determinants instead. In realistic applications, the associated exponential cost can be avoided through the introduction of explicit particle-number projection operators \cite{Fanto:2017ei,Ryssens:2021nq,Lang:1993dl}. For the applications to schematic models that are considered in this work, we explicitly work with a fixed-A Slater determinant basis, since we will exactly diagonalize the Hamiltonians for benchmarking in any case.

Using Boltzmann factors, the probability that the system is in the many-body state where the energy levels $\epsilon_{p_1}, \epsilon_{p_2},..., \epsilon_{p_A}$ are occupied is 
\begin{equation}
    P(\{p_1,\ldots,p_A\}) = \frac{e^{-(\epsilon_{p_1}+\epsilon_{p_2}+...+\epsilon_{p_A})\beta}}{Z_A},
\end{equation}
and the occupation number $n_p$ is the sum of these probabilities for all many-body states in which the energy level $\epsilon_p$ is occupied. Adopting the notation of \cite{Barghathi:2020zn}, this can be conveniently written as
\begin{equation}
    n_p = \frac{Z^{\setminus \{p\}}_{A-1}e^{-\beta \epsilon_p}}{Z_A}, \label{eq:ce_occ}
\end{equation}
where $Z^{\setminus \{p\}}_{A-1}$ is the partition function for $A-1$ fermions where the energy level $\epsilon_p$ is removed. 

The canonical FT-HF single-particle energies and single-particle wave functions can now be found by minimizing the (Helmholtz) free energy
\begin{equation}\label{helm}
    F = E - TS\,.
\end{equation}
The definitions of $\epsilon_p$ and $E$ Eqs.~\eqref{eq:gce_fthf_eps} and \eqref{eq:gce_fthf_energy} are unchanged, but are now understood in terms of the canonical occupation numbers \eqref{eq:ce_occ}. Thus, the partition function for canonical FT-HF is given as
\begin{equation} 
  Z^\text{HF}_A = e^{\beta V_\text{HF}}
      \sum^{m_1+\ldots m_N = A}_{\{m_i\}} \bra{\{m_i\}} e^{-\beta\epsilon_p\,a^\dagger_pa_p} \ket{\{m_i\}}\,.\label{eq:ce_fthf_partition}
\end{equation}
The canonical entropy can be computed as 
\begin{equation}
    S^\text{HF}_A = \ln Z^\text{HF}_A + \beta E \label{eq:ce_fthf_entropy}
\end{equation}
(see, e.g., \cite{Ryssens:2021nq}).

We conclude this discussion by describing our implementation of the FT-HF iteration procedure. We begin by expressing the Hamiltonian as
\begin{equation}\label{preham}
    \hat{H} = \hat{K} + \hat{V} = \sum_{\alpha\beta} k_{\alpha\beta} a^\dagger_\alpha a_\beta + \frac{1}{4} \sum_{\alpha\beta\gamma\delta} v_{\alpha\beta\gamma\delta} a^\dagger_\alpha a^\dagger_\beta a_\delta a_\gamma\,,
\end{equation}
where we use Greek indices to indicate a general working basis of single-particle states.
We will then define the effective HF Hamiltonian $\hat{f}$, starting with the ansatz
\begin{equation}
    \hat{f}^{(0)} = \hat{k}.
\end{equation}
We diagonalize $\hat{f}$, yielding eigenvalues $\{ \epsilon_p \}$ and eigenstates that define a unitary similarity transformation $\hat{u}$. 

Next, we calculate the grand-canonical or canonical occupation numbers that minimize the grand potential or free energy, respectively, assuming $A$ independent fermions with energy levels $\{ \epsilon_p \}$. 

Using the occupation numbers $n_p$ and the previously mentioned unitary transformation, we iteratively construct the one-body density matrix
\begin{equation}\label{hfrho}
    \rho_{\alpha\beta}^{(i)} = \sum_p n_p^{(i)}u_{\alpha p}^{(i)}u_{\beta p}^{(i)}\,,
\end{equation}
as well as a new $\hat{f}$ and HF energy $E$ using
\begin{align}\label{hff}
    f_{\alpha\beta}^{(i+1)} &= k_{\alpha\beta} + \sum_{\gamma\delta} v_{\alpha\gamma\beta\delta} \rho_{\gamma\delta}^{(i)} \\
    \label{hfe} E^{(i+1)} &= \sum_{\alpha\beta} k_{\alpha\beta} \rho_{\alpha\beta}^{(i)} + \frac{1}{2}\sum_{\alpha\beta\gamma\delta} v_{\alpha\beta\gamma\delta} \rho_{\alpha\gamma}^{(i)} \rho_{\beta\delta}^{(i)}.
\end{align}
Note that in the case of a diagonal density matrix (which will occur if $\hat{f}$ is diagonal), these equations resemble the normal ordering of Eqs. \eqref{noe} and \eqref{nof}. 

We summarize the process below:
\begin{enumerate}
    \item Diagonalize $\hat{f}$, yielding eigenvalues $\{ \epsilon_p \}$ and a unitary transformation $\hat{u}$.
    \item Calculate the (grand-)canonical occupation numbers $n_p$ based on the eigenvalues $\{ \epsilon_p \}$.
    \item Construct the density matrix via Eq. \eqref{hfrho}.
    \item Construct a new $\hat{f}$ and HF energy $E$ using Eqs. \eqref{hff} and \eqref{hfe}.
\end{enumerate}
We repeat these steps until the energy $E$ is converged, which we define to be the case once 
\begin{equation}
    \frac{\left| E^{(i+1)} - E^{(i)} \right|}{A} \leq 10^{-5}.    
\end{equation}
Checking the convergence of $E$ instead of $F$ or $\Omega$ is justified because it can be shown that the FT-HF minima of these quantities coincide for fixed thermal occupation numbers, see \cite{Sanyal:1993hh,Sanyal:1994ly}.

Finally, we transform the Hamiltonian (Eq. \eqref{preham}) to the FT-HF basis (Eq. \eqref{hf-ham}), using the eigenvalues $\{n_p\}$ and eigenstates $\{U_p\}$ of the final density matrix:
\begin{equation}
    K_{pq} = \sum_{\alpha\beta} U^*_{p\alpha}k_{\alpha\beta}U_{\beta q}.
\end{equation}
and
\begin{equation}
    V_{pqrs} = \sum_{\alpha\beta\gamma\delta} U^*_{p\alpha}U^*_{q\beta}v_{\alpha\beta\gamma\delta}U_{\gamma r}U_{\delta s}.
\end{equation}

\subsection{The FT-IMSRG Flow} \label{sec3b}
Let us now consider the implementation of an IMSRG flow at finite temperature. Since finite-temperature observables of interest are defined as traces over many-body configurations, decoupling a single state from the others like we did in the zero-temperature IMSRG is no longer very useful. Note, however, that the same IMSRG transformation that decoupled the reference state also eliminates couplings that change the particle-hole excitations of a basis configuration by $\pm 2$ (up to truncation errors), as indicated by the elimination of the side-diagonals in Fig.~\ref{fig:decoupling}. As discussed in Sec.~\ref{sec2}, this implicit reshuffling of correlations will be our guiding principle for the finite-temperature case as we will seek to construct an RG-improved Hamiltonian for which the evaluation of thermal expectation values is (greatly) simplified.

Let us first consider the grand canonical ensemble. The density operator and partition function are given by Eqs.~\eqref{eq:gce_def_rho} and \eqref{eq:gce_def_z}, respectively.
We can partition the Hamiltonian as $\hat{H}=\hat{H}_0 + \hat{H}_1$, such that the ensemble consists of eigenstates of $\hat{H}_0$, and $\hat{H}_1$ contains all remaining interactions. Introducing a thermal interaction picture with operators $\hat{O}(\beta) = e^{\beta(\hat{H}_0-\mu\hat{A})} \hat{O} e^{-\beta(\hat{H}_0-\mu\hat{A})}$, the partition function can be rewritten as \cite{Fetter:2003ve} 
\begin{equation}\label{eq:partition}
    Z = Z_0\;\Big\langle\mathcal{T} \exp \int^\beta_0 d\tau\; (-\hat{H}_1(\tau))\Big\rangle\,,
\end{equation}
where $\mathcal{T}$ is a path-ordering operator, $Z_0 = \Tr e^{-\beta(\hat{H}_0-\mu \hat{N})}$ and the expectation value is taken with respect to $\hat{\rho}_0$:
\begin{equation}
    \expect{\hat{O}} \equiv \Tr \left( \hat{\rho}_0\hat{O}\right)
    = \frac{1}{Z_0} \Tr\left( e^{-\beta(\hat{H}_0-\mu \hat{A})}\hat{O}\right)\,.
\end{equation}
We obtain analogous expressions when working with the canonical ensemble by starting from $\rho_A$ and $Z_A$ \eqref{eq:ce_def_rho} and partitioning the Hamiltonian in the same way. The path-ordered exponential in Eq.~\eqref{eq:partition} can be treated with systematic perturbative or non-perturbative expansion methods \cite{Fetter:2003ve,Santra:2017tk,Hirata:2020bf,Hirata:2021lj,White:2018sp,Harsha:2019qi}. In our case, $Z_0$ and $\hat{H}_0$ will be given by FT-HF and the latter will only consist of zero- and one-body operators, as discussed in the previous section.

Using the grand canonical FT-HF partition function and density operator, the normal ordering and Wick's theorem used in Sec.~\ref{sec2} generalize to finite temperature in a straightforward way \cite{Gaudin:1960ud,Fetter:2003ve,Schonhammer:2017ms}. One-body density matrices and occupation numbers simply need to be replaced by their finite-temperature counterparts: For example,
\begin{align}
    \contraction[1.5ex]{}{a}{a}{}a^\dagger_pa_q &\equiv \expect{a^\dagger_p a_q} = n_p \delta_{pq} \,,\label{eq:wicka}\\
    \contraction[1.5ex]{}{a}{a}{}a_pa^\dagger_q &\equiv \expect{a_p a^\dagger_q} = \overline{n} \delta_{pq} \,,\label{eq:wickb}
\end{align}
and
\begin{align}
    \expect{a^\dagger_p a^\dagger_q a_s a_r} &= \expect{a^\dagger_pa_r}\expect{a^\dagger_qa_s} -
    \expect{a^\dagger_pa_s}\expect{a^\dagger_qa_r}\notag\\
    &= n_p n_q \left(\delta_{pr}\delta_{qs} - \delta_{ps}\delta_{qr}\right)\,,\label{eq:wickc}
\end{align}
with the average thermal occupation numbers defined by Eqs. \eqref{eq:gce_occ} or \eqref{eq:ce_occ}. 

As discussed in Sec.~\ref{sec3a}, the canonical ensemble is a more appropriate choice for a finite system like a nucleus. Even though the canonical FT-HF Hamiltonian and partition function still only involve up to one-body operators (cf. Eqs.~\eqref{eq:gce_fthf_eps} and \eqref{eq:ce_fthf_partition}), Wick's theorem must be amended with additional terms that ensure that the particle number is fixed \cite{Schonhammer:2017ms}. In our present implementation, the evalution of these corrections would incur exponential computational cost beyond the computation of the canonical average occupation numbers (cf. Sec~\ref{sec3a}), hence we neglect them in the normal ordering and the FT-IMSRG flow equations, and only include the standard contractions \eqref{eq:wicka}, \eqref{eq:wickb}, but with the canonical occupation numbers \eqref{eq:ce_occ}. Our results in Section~\ref{sec4} justify this procedure as a reasonable approximation.

Since explicit particle-number projection operators can be used to switch from the grand canonical to the canonical ensemble (cf.~Sec.~\ref{sec3a} and Refs.~\cite{Fanto:2017ei,Ryssens:2021nq}, we expect that the correction terms defined in  \cite{Schonhammer:2017ms} can be obtained in a different way through the generalized normal-ordering formalism and Wick's theorem of Mukherjee and Kutzelnigg \cite{Kutzelnigg:1997fk}, which is designed for arbitrary wave functions and ensembles. This framework has been used to implement zero-temperature multi-reference IMSRG with particle-number projected Hartree-Fock Bogoliubov reference states  \cite{Hergert:2017kx,Hergert:2020am}. Consequently, the MR-IMSRG code platform already contains many ingredients for grand canonical and fully canonical FT-IMSRG calculations with realistic nuclear Hamiltonians, and it will be straightforward to extend it to finite temperature in the near future.

Next, we need to construct an appropriate generator for the FT-IMSRG. In the zero-temperature case, we identified $f_{ai}$ and $\Gamma_{abij}$ (and their Hermitian conjugates) as the components of the Hamiltonian that should be eliminated to decouple the reference state. These particular coefficients also appear in all perturbative corrections to the ground-state energy and other observables, since they are responsible for the initial particle-hole excitation of the ground state and/or the eventual de-excitation back to that state \cite{Hergert:2016jk,Zare:2023lg}. As they are driven to zero by the IMSRG, the associated correlations are directly built into the renormalized Hamiltonian. 

This view of the IMSRG transformation readily generalizes to the finite-temperature setting: We seek to achieve
\begin{equation}
     \Tr\left(\hat{\rho}\hat{O}\right)\;\xrightarrow[s\to\infty]\qquad \Tr\left(\hat{\rho}_0\hat{O}(\infty)\right) = \expect{\hat{O}(\infty)}\,.
\end{equation}
This is prompting us to suppress all expectation values of the form
\begin{equation}\label{1b-decoupling}
    \expect{\hat{H} \{a^\dagger_p a_q\}} = \n_pn_qf_{pq}
\end{equation}
and
\begin{equation}\label{2b-decoupling}
    \expect{\hat{H} \{a^\dagger_p a^\dagger_q a_s a_r\}} = \n_p\n_qn_rn_s\Gamma_{pqrs},
\end{equation}
in analogy to the off-diagonal matrix elements \eqref{eq:f_od},\eqref{eq:Gamma_od} of the zero-temperature formalism \cite{Santra:2017tk,Hirata:2021lj}. 

Using these matrix elements, we construct generators of the form
\begin{align}
    \hat{\eta} &= \sum_{pq} \n_pn_q\eta_{pq}\{a^\dagger_p a_q\}  \notag\\ &\hphantom{=}+\frac{1}{4} \sum_{pqrs} \n_p\n_qn_rn_s\eta_{pqrs}\{a^\dagger_p a^\dagger_{q} a_{s} a_r\} - \textrm{H.c.}
\end{align}
The energy denominators used in our generators will become 
\begin{align}
    \Delta_{pq} &= \expect{\{a^\dagger_q a_p\}\hat{H}\{a^\dagger_p a_q\}} - \expect{ \hat{H}} \notag \\
    &= \n_pn_q\left(\n_pf_{pp} - n_qf_{qq} - \n_pn_q\Gamma_{pqpq}\right)
\end{align}
and
\begin{align}
    \Delta_{pqrs} &= \expect{\{a^\dagger_r a^\dagger_s a_q a_p\}\hat{H}\{a^\dagger_p a^\dagger_q a_s a_r\}} - \expect{\hat{H}} \notag \\
    &= \n_p\n_qn_rn_s\bigg(\n_p f_{pp} + \n_q f_{qq} - n_r f_{rr} - n_s f_{ss} \notag \\ &\qquad + \n_p\n_q\Gamma_{pqpq} + n_rn_s\Gamma_{rsrs} - \n_pn_r\Gamma_{prpr} \notag \\ &\qquad - \n_qn_s\Gamma_{qsqs} - \n_pn_s\Gamma_{psps} - \n_qn_r\Gamma_{qrqr} \bigg).
\end{align}

We also insist that in the zero-temperature limit (where the occupation numbers become 0 and 1), our generator approaches a zero-temperature generator. If we naively attempt to generalize the White generator to finite-temperature, we find that due to the occupation factors $\n_pn_q$ and $\n_p\n_qn_rn_s$ appearing in both the numerators and denominators, taking the zero-temperature limit will leave us with many nonzero and potentially divergent generator matrix elements for non-ph or -pphh entries. To address this problematic behavior, we use the arctan variant of the generator
\begin{align}
    \hat{\eta} &= \frac{1}{2} \sum_{pq} \n_pn_q \arctan\left(\frac{2f_{pq}}{\Delta_{pq}}\right) \{a^\dagger_p a_q\} \notag \\ & + \frac{1}{8} \sum_{pqrs} \n_p\n_qn_rn_s \arctan \left(\frac{2\Gamma_{pqrs}}{\Delta_{pqrs}}\right)\{a^\dagger_p a^\dagger_q a_s a_r\} \notag \\ &\qquad - \textrm{H.c.} \label{eq:white_atan_ft}
\end{align}
which does behave properly in the zero-temperature limit. It also regularizes potential singularities stemming from vanishing $\Delta_{pq}$ or $\Delta_{pqrs}$, although we did not encounter any in the applications discussed in the present work.

Before we proceed, a few additional comments are in order. First, we note 
that the computational scaling of the FT-IMSRG is returned to $\mathcal{O}(N^6)$, because we are no longer able to distinguish between particles and holes. Second, closer inspection of the generator \eqref{eq:white_atan_ft} shows that it does not directly drive Eqs.~\eqref{1b-decoupling} and \eqref{2b-decoupling} to zero. Splitting these expectation values into commutator and anticommutator parts, 
\begin{align}
    \expect{\hat{H}(\infty)\{a_p^\dagger\ldots a_q\}}
    &= \frac{1}{2}\expect{\left[\hat{H}(\infty),\{a_p^\dagger\ldots a_q\}\right]_+}\notag\\
    &\hphantom{=}+ \frac{1}{2}\expect{\left[\hat{H}(\infty),\{a_p^\dagger \ldots a_q\}\right]}\,,
\end{align}
we see that only the latter is guaranteed to vanish as we evolve $s\to\infty$, although the former may be numerically reduced in size:
\begin{align}
    \expect{\left[\hat{H}(\infty),\{a_p^\dagger a_q\}\right]} = 0\,, \label{eq:ibc1} \\
    \expect{\left[\hat{H}(\infty),\{a_p^\dagger a_q^\dagger a_s a_r\}\right]} = 0\,.\label{eq:ibc2}
\end{align}
As discussed in detail in Ref.~\cite{Hergert:2017kx,Hergert:2020am}, this is the case for all of the usual IMSRG generators. It is a direct consequence of working with a unitary transformation, while implementing Eqs.~\eqref{1b-decoupling} and \eqref{2b-decoupling} would require a more general transformation \cite{Mukherjee:2001uq}. This issue previously arose in the zero-temperature MR-IMSRG due to the use of correlated reference states, but it does not appear in the standard zero-temperature IMSRG discussed in Sec. \ref{sec2}, because it is easy to see that the distinct particle or hole nature of the single-particle states immediately results in
\begin{align}
    &\expect{\hat{H}\{a_p^\dagger \ldots a_q\}} \notag\\ 
    &=\expect{\left[\hat{H},\{a_p^\dagger \ldots a_q\}\right]_+} = 
    \expect{\left[\hat{H},\{a_p^\dagger\ldots a_q\}\right]}
\end{align}
at arbitrary values of the flow parameter $s$.

The conditions \eqref{eq:ibc1} and \eqref{eq:ibc2} are known as the irreducible one- and two-body Brillouin conditions, IBC(1) and IBC(2). 
As discussed in Refs. \cite{Kutzelnigg:1980ci,Mukherjee:2001uq,Mazziotti:2006fk}, they define a systematic many-body truncation hierarchy.
Geometrically, the expectation values of the commutators are gradients of 
$\expect{\hat{H}}$ under unitary variations. Thus, $\expect{\hat{H}(\infty)}$ will be extremal --- usually minimal --- when $\hat{\eta}$ vanishes at the fixed point of the FT-IMSRG evolution. To make a connection with our previous discussion about the reshuffling of correlations, we also note that (at least) the second-order FT-MBPT energy correction can be rewritten in terms of the commutators appearing in the IBC(1) and IBC(2), so Eqs.~\eqref{eq:ibc1} and \eqref{eq:ibc2} imply that this correction vanishes for the evolved Hamiltonian $\hat{H}(\infty)$.

\subsection{Particle Number Variance}
One of the complications of using a thermal ensemble as opposed to a single Slater determinant reference state is that the ensemble can run over states with different particle numbers, and the particle number variance $\Delta A = \langle \hat{A}^2 \rangle - \langle \hat{A} \rangle ^2$ may be nonzero. 

We have
\begin{equation}
    \hat{A} = \sum_{p} a_p^\dagger a_p
\end{equation}
and thus
\begin{align}
    \hat{A}^2 &= \sum_{pq} a_p^\dagger a_p a_q^\dagger a_q = \sum_{pq} \left( a_p^\dagger a_q^\dagger a_q a_p + \delta_{pq}a_p^\dagger a_q \right)\notag\\
    &= \sum_{pq} a_p^\dagger a_q^\dagger a_q a_p + \sum_p a_p^\dagger a_p.
\end{align}
We can then normal order the above operators:
\begin{equation}\label{ano}
    \hat{A} = \sum_p n_p + \sum_{p} \{ a_p^\dagger a_p\} = A + \sum_{p} \{ a_p^\dagger a_p\}
\end{equation}
and
\begin{align}
    \hat{A}^2 &= \sum_p (n_p - n_p^2) + \sum_{pq}n_pn_q + \sum_{p} \{ a_p^\dagger a_p\}  \notag\\
    &\qquad +\sum_{pq} \{a_p^\dagger a_q^\dagger a_q a_p\} \\
    &= \sum_p \n_p n_p + A^2 + \sum_{p} \{ a_p^\dagger a_p\} + \sum_{pq} \{a_p^\dagger a_q^\dagger a_q a_p\}.
\end{align}
This means that
\begin{equation}
    \langle \hat{A} \rangle = A
\end{equation}
and
\begin{equation}
    \langle \hat{A}^2 \rangle = A^2 + \sum_p \n_p n_p,
\end{equation}
so that
\begin{equation}
    \Delta A = \langle \hat{A}^2 \rangle - \langle \hat{A} \rangle ^2 = \sum_p \n_p n_p.
\end{equation}

While it appears that this result is in full generality, the canonical ensemble is defined such that $\Delta A = 0$. The previously-mentioned corrections to Wick's theorem in the canonical ensemble \cite{Schonhammer:2017ms} enforce the condition $\Delta A = 0$. As explained above, we neglect them in our present implementation because because of their scaling with the exponential size of the many-body basis, but they can be included in future realistic calculations through the use of explicit particle-number projection techniques (see Sec.~\ref{sec3b} and Refs.~\cite{Fanto:2017ei,Ryssens:2021nq,Hergert:2017kx,Hergert:2020am}. 

Finally, we note that Eq. \eqref{ano} holds in both ensembles, and we can use it to determine the IMSRG flow of $\hat{A}$ by calculating $[\hat{\eta}, \hat{A}]$. At $s=0$, the zero-body part is
\begin{equation}
    [\hat{\eta}, \hat{A}]^{0B} = \sum_{pq}(n_p-n_q)\eta_{pq} \delta_{pq} = 0, \label{eq:A_0B}
\end{equation}
the one-body part is
\begin{align}
    [\hat{\eta}, \hat{A}]^{1B}_{pq} &= \sum_r (1+P_{pq})\eta_{pr}\delta_{rq} - \sum_{rs} (n_r-n_s)\delta_{rs}\eta_{sprq} \notag\\
    &= \eta_{pq} + \eta_{qp} = 0 \label{eq:A_1B}
\end{align}
due to the anti-Hermiticity of $\hat{\eta}$, and the two-body part is
\begin{align}
    [\hat{\eta}, \hat{A}]^{2B}_{pqrs} &= -\sum_t \left( (1-P_{pq})\delta_{pt}\eta_{tqrs} - (1-P_{rs})\delta_{tr}\eta_{pqts} \right) \notag\\
    &= - \left(\eta_{pqrs} - \eta_{qprs} + \eta_{pqrs} - \eta_{pqsr} \right) = 0. \label{eq:A_2B}
\end{align}
This implies that the there is no change to $\hat{A}$ from the initial derivative at $s=0$, especially no induced two-body (or higher-rank) contribution. This also means that $\hat{A}$ remains a pure one-body operator through the evolution, so Eqs.~\eqref{eq:A_0B}--\eqref{eq:A_2B} will be valid for any $s$ and we see that
\begin{equation}
    \frac{d}{ds}\hat{A} = [\hat{\eta}(s), \hat{A}] = 0\,.
\end{equation}
Using this result, it is easy to prove that
\begin{equation}
    [\hat{\eta}, \hat{A}^2] = \hat{A}[\hat{\eta}, \hat{A}] + [\hat{\eta}, \hat{A}] \hat{A} =  0.
\end{equation}
Thus, the average particle number as well as the particle number variance are invariant under the FT-IMSRG flow, and \emph{entirely} determined by the reference ensemble.

\section{Results} \label{sec4}
\subsection{The P3H Hamiltonian} \label{sec4a}
To test the FT-IMSRG, we employ the pairing-plus-particle-hole (P3H) model, which is exactly solvable and qualitatively captures important features of nuclear interactions \cite{Hjorth-Jensen:2010qf}. We work with single-particle states $\{(\alpha, \sigma)\}$ where $\alpha=1,2,3,...,N/2$ is the principal quantum number and $\sigma=+,-$ represents the spin. We do not make any assumptions about the form of the single-particle wave functions, and we do not impose any symmetries in our calculations.

The P3H Hamiltonian is governed by the parameters $\delta$, $g$, and $b$, and, in the notation of Eq. \eqref{preham}, has the one- and two-body parts
\begin{equation}
    \hat{K}_\textrm{P3H} = \delta \sum_{\alpha\sigma} (\alpha-1) a^\dagger_{\alpha\sigma} a_{\alpha\sigma}
\end{equation}
and
\begin{align}
    \hat{V}_\textrm{P3H} &= -\frac{g}{2}\sum_{\alpha\beta} a^\dagger_{\alpha+} a^\dagger_{\alpha-} a_{\beta-} a_{\beta+} \notag \\ 
    &\quad - \frac{b}{2}\sum_{\alpha,\beta,\gamma\neq \beta} \bigg( a^\dagger_{\alpha+} a^\dagger_{\alpha-} a_{\beta-} a_{\gamma+} \notag \\ &\quad\qquad\qquad\qquad+ a^\dagger_{\gamma+} a^\dagger_{\beta-} a_{\alpha-} a_{\alpha+}\bigg).
\end{align}
Here, $\delta$ is the (constant) spacing between single-particle energy levels, $g$ is the strength of pairing interaction, and $b$ controls the strength of pair-breaking, particle-hole type excitations. The structure of the eigenstates will be driven by the competition between the pairing and pair-breaking interactions, as well as the ratios $g/\delta$ and $b/\delta$, i.e., the ability of the interaction terms to overcome the level spacing. Because of the latter observation, for the remainder of this paper we express all quantities in natural units defined by $\delta=1$, without loss of generality.

\subsection{Four Particles in Eight States} \label{sec4b}
We first consider the case of four fermions in eight possible single-particle states (i.e., $A=4$, $N=8$). The many-body basis will have dimension 70, so the exact solution can be easily computed for comparison. Note that for this section, occupation numbers will be computed in the canonical ensemble so that the FT-IMSRG can be tested with the most accurate FT-HF input possible.

\begin{figure}[t]
    \centering
    \includegraphics[width=0.45\textwidth]{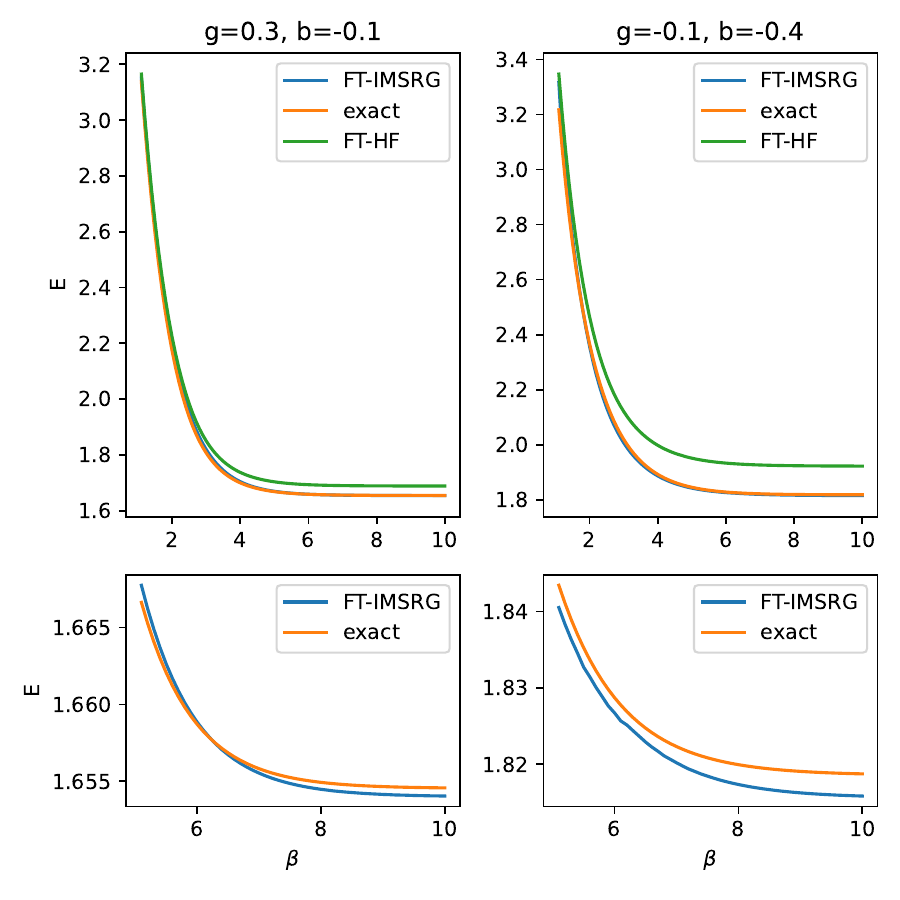}
    \caption{Internal energy versus inverse temperature with $A=4$ and $N=8$ for $g=0.3$, $b=-0.1$ (left) and $g=-0.1$, $b=-0.4$ (right). The bottom panels show the low-temperature range, $\beta\geq 5.0$. The FT-IMSRG results (blue) are much closer to the exact results (orange) than the FT-HF results (green) are. The FT-IMSRG results successfully replicate the behavior of the exact results at low temperatures.}
    \label{fig:energyA4N8}
\end{figure}

Fig. \ref{fig:energyA4N8} plots $E$ versus $\beta$ for different coupling strengths. In general, the FT-IMSRG significantly improves the FT-HF results in accuracy. Particularly at low temperatures (high $\beta$), the FT-IMSRG results are extremely close to the exact internal energies.

\begin{figure}[t]
    \centering
    \includegraphics[width=0.45\textwidth]{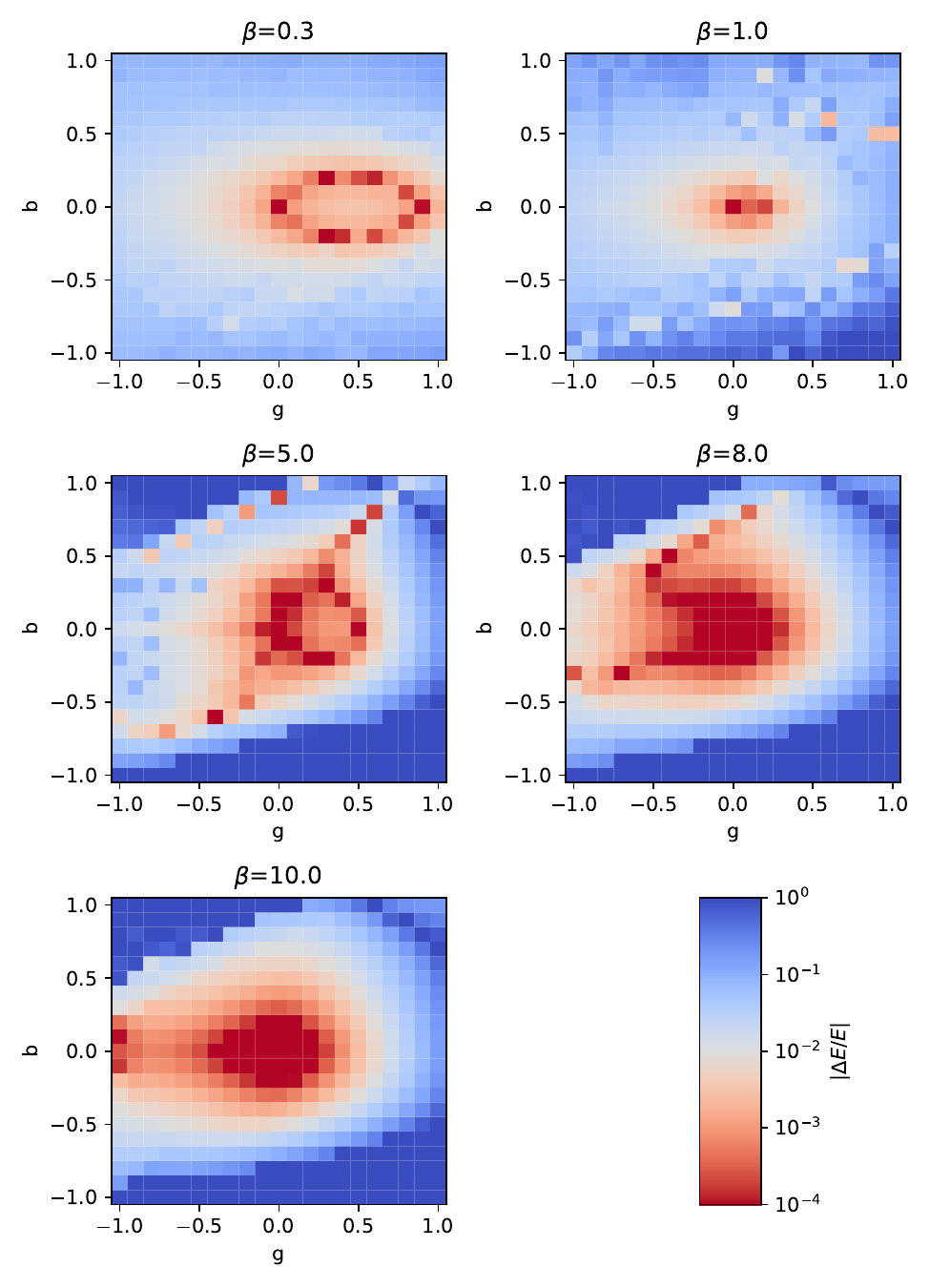}
    \caption{Relative error in internal energy in parameter space for various inverse temperatures with a logarithmic color scale. Shades of red denote less than 1\% error. The FT-IMSRG results display the most accuracy for lower temperatures and parameters of lower magnitude.}
    \label{fig:cplotA4N8}
\end{figure}

As would be expected, the FT-IMSRG performs best for weak coupling in the Hamiltonian (i.e. low $|g|$ and $|b|$). For $|g|, |b|\leq 0.5$, the FT-IMSRG results consistently demonstrate good agreement with the exact results. This can be seen in Fig. \ref{fig:cplotA4N8}. At lower temperature, we see strong agreement for the widest range of parameters. The performance of the FT-IMSRG is weakest at mid-range temperatures, around $\beta = 1$, and is improved as $\beta \to 0$. This is because at these temperatures a wider range of single-particle states have significant nonzero occupation numbers, and thus the decoupling conditions \ref{1b-decoupling} and \ref{2b-decoupling} become more demanding. However, as $\beta\to 0$, the thermal energy becomes much greater than the interaction strength, and we approach the limit of a free theory. While the FT-IMSRG still improves upon the FT-HF results for mid-range temperatures, to achieve a similar accuracy to that achieved at lower temperatures further improvements to the FT-IMSRG truncation scheme would be necessary (see, e.g., \cite{Heinz:2021mk,He:2024vl,Stroberg:2024cr} for recent discussions).

When $|g|$ and $|b|$ become too large, specifically when they are of opposite sign, the FT-IMSRG frequently diverges (shown in deep blue in Fig. \ref{fig:cplotA4N8}). This is because positive $g$ encourages pairing, and negative $b$ discourages pair-breaking, and vice versa, leading to a mutual reinforcement.

\begin{figure}[t]
    \centering
    \includegraphics[width=0.45\textwidth]{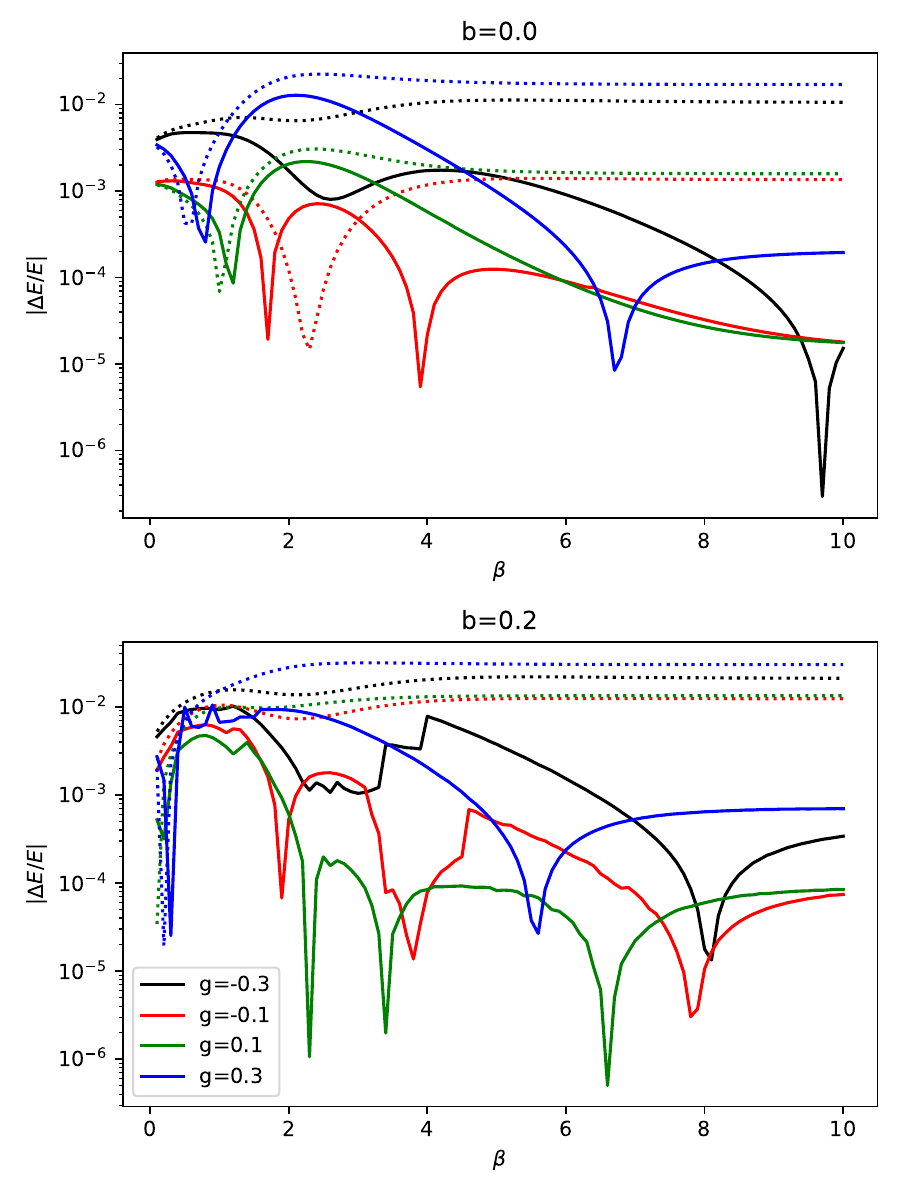}
    \caption{Relative error in internal energy versus inverse temperature for $b=0$ (top) and $b=0.2$ (bottom) with various values of $g$ on a logarithmic scale. The FT-HF results are shown with dotted lines and the FT-IMSRG results are shown with solid lines. In all cases, the FT-IMSRG results show a significant improvement compared to the FT-HF results, with better results for weaker coupling.}
    \label{fig:logerrorA4N8}
\end{figure}

The relative error of the internal energy as a function of $\beta$ is shown on a logarithmic plot for various coupling strengths in Fig. \ref{fig:logerrorA4N8}. Once again, we see the strong agreement between the FT-IMSRG and the exact results, which is strongest with weaker coupling and lower temperatures. The less smooth results in the bottom panel are likely due to numerical complications that arise from the pair-breaking term of the Hamiltonian. The relative errors seen here are comparable to those of Finite-Temperature Coupled Cluster for similar schematic models (see, e.g., \cite{White:2020dn,Harsha:2022dw}). 

\begin{figure}[t]
    \centering
    \includegraphics[width=0.45\textwidth]{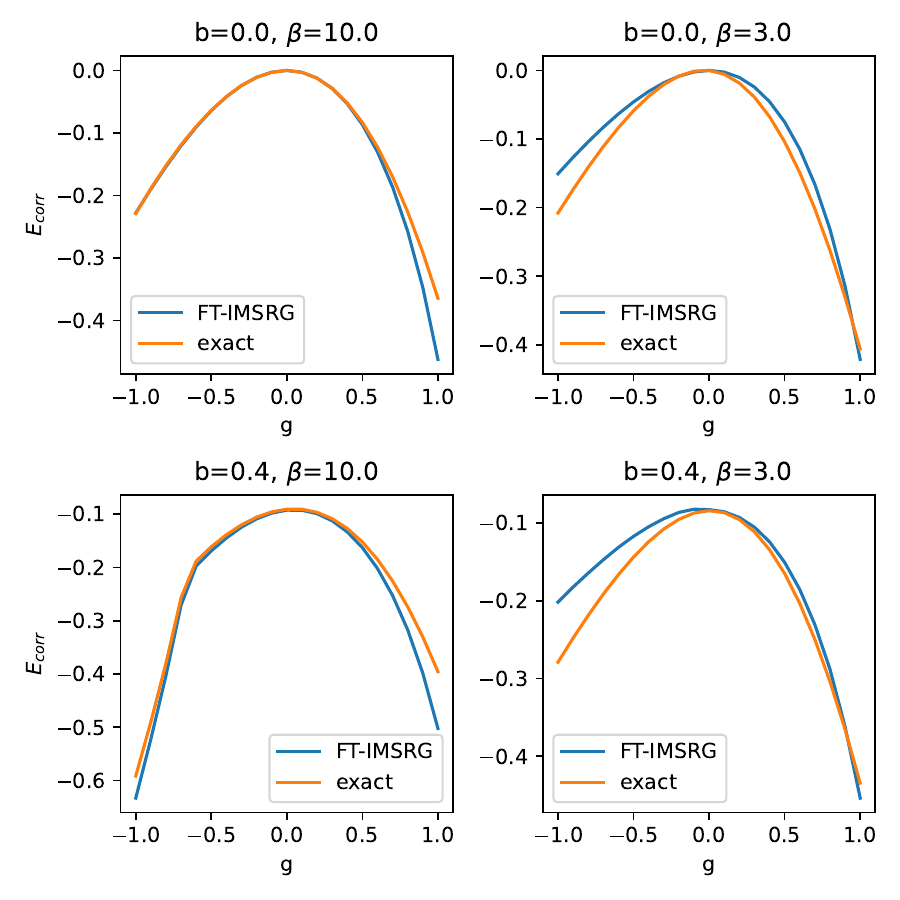}
    \caption{Correlation energy versus $g$ for various values of $b$ and $\beta$. At low temperature, the FT-IMSRG results (blue) and exact results (orange) are in great agreement. Weaker coupling in both $b$ and $g$ produce more accurate FT-IMSRG results.}
    \label{fig:corrA4N8}
\end{figure}

We also show plots of the correlation energy
\begin{equation}
    E_\textrm{corr} = E - E_\mathrm{HF}
\end{equation}
versus the pairing strength $g$ for different $\beta$ and $b$ values in Fig. \ref{fig:corrA4N8}. Interestingly, at higher temperatures, more deviations are observed for negative $g$ (attractive pairing), and at lower temperatures, more deviations are observed for positive $g$ (repulsive pairing). This trend is apparent in Fig. \ref{fig:cplotA4N8} as well. The behavior at low temperature can be explained from the fact that the IMSRG(2) is known to under count a subset of fourth-order perturbation theory contributions by a factor of $1/2$, see~\cite{Stroberg:2024cr,Hergert:2016jk}. Higher temperatures weaken the effects of heavily-favored pairing, but exacerbate the effects of heavily-disfavored pairing so that the interaction cannot be fully accounted for by the IMSRG. This leads to the behavior we observe in the correlation energies as temperature increases.

\begin{figure}[t]
    \centering
    \includegraphics[width=0.45\textwidth]{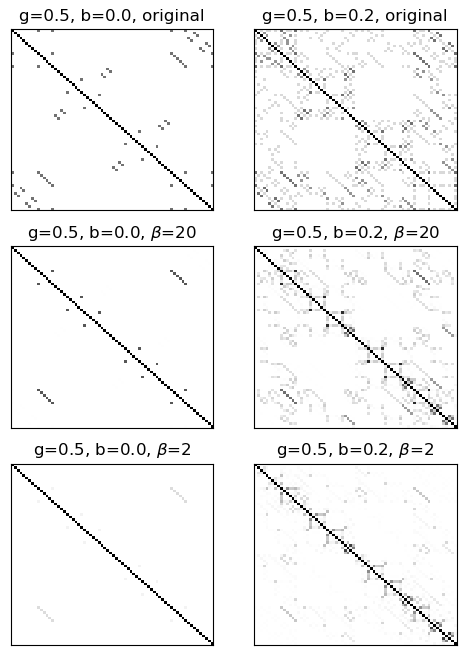}
    \caption{The full 70$\times$70 many-body Hamiltonian matrix for $A=4$ and $N=8$ and two sets of parameters. We compare the unevolved Hamiltonian at zero temperature ($\beta\to\infty$, row), to the FT-IMSRG evolved Hamiltonian for $\beta=20$ (middle row), and for $\beta=2$ (bottom row). Darker colors correspond to matrix elements with larger absolute values, and matrix elements with a value of zero are shown in white. At low temperatures, the FT-IMSRG decouples only the lowest-energy states, while at higher temperatures the FT-IMSRG decouples many more states.}
    \label{fig:matrices}
\end{figure}

It is also informative to look at how the FT-IMSRG decouples the many-body Hamiltonian matrix at different temperatures (recall the decoupling behavior of the zero-temperature IMSRG shown in Fig. \ref{fig:decoupling}). In Fig. \ref{fig:matrices}, we show this for the example of a pure pairing Hamiltonian with $g=0.5$, as well as the same Hamiltonian when a pair-breaking term is added with $b=0.2$. The normal-ordered pieces of the Hamiltonian after the FT-IMSRG flow are de-normal ordered and used to build the full many-body Hamiltonian matrix. At low temperatures, the FT-IMSRG decouples only the lowest-energy states as is the case in the zero-temperature IMSRG, while at higher temperatures the FT-IMSRG decouples many more states. However, this comes at the cost of truncation errors. After computing the eigenvalues of these matrices, we compare the exact free energies. At $\beta=20$ (very low temperature), the truncation error is under 0.25\% for the pure pairing Hamiltonian, and under 0.55\% once the pair-breaking interaction is added. At higher temperature, with $\beta=2$, the truncation error is about 0.86\% for the pure pairing Hamiltonian, but grows to a little less than 3\% once the pair-breaking interaction is added. These truncation errors can help explain the decrease in the FT-IMSRG's accuracy at high temperatures.

\subsection{Increasing Particle Number and Basis Size} \label{sec4c}

\begin{figure}[t]
    \centering
    \includegraphics[width=0.45\textwidth]{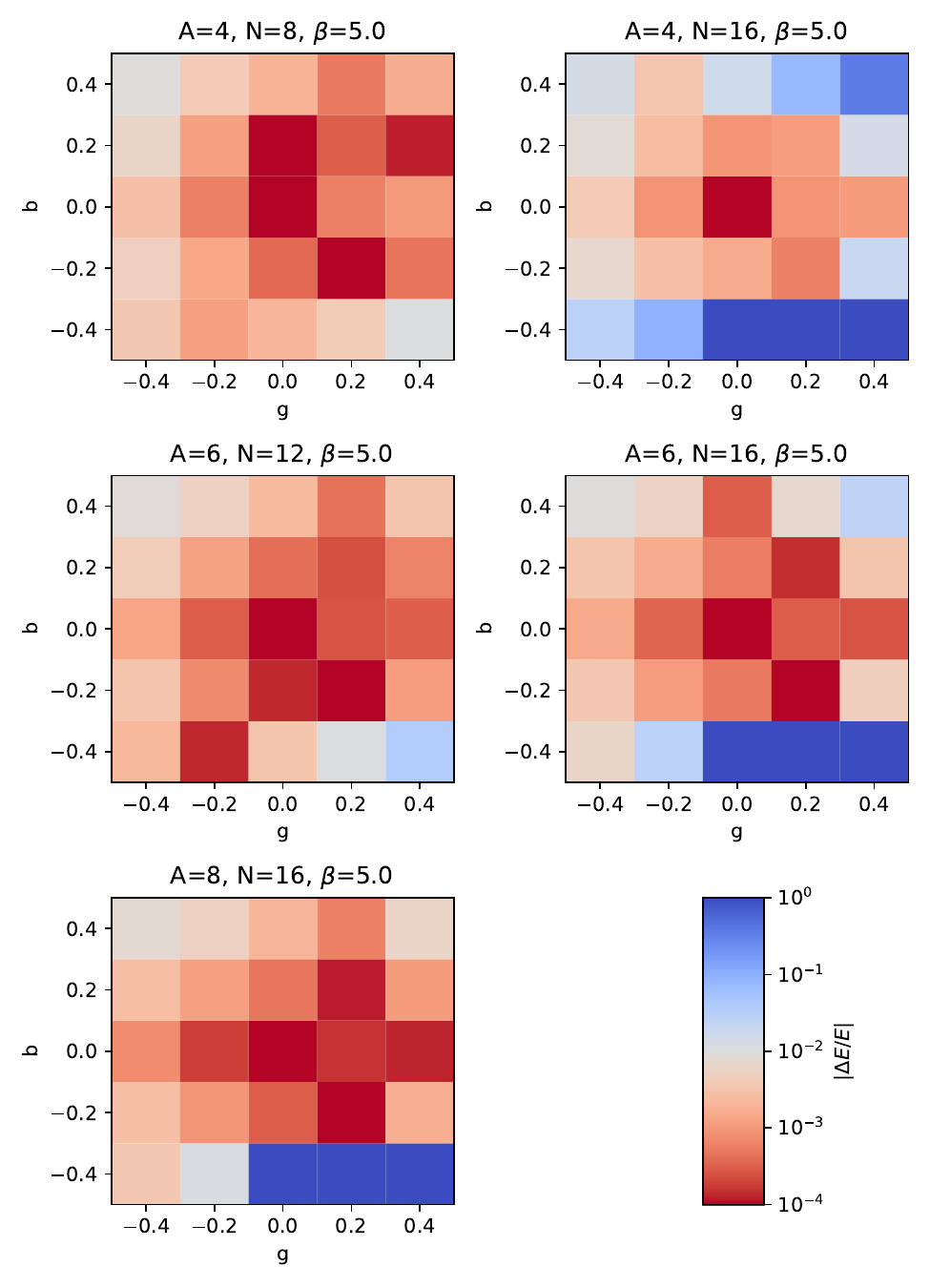}
    \caption{Relative error in internal energy in parameter space for $\beta=5.0$ and various values of $A$ and $N$ with a logarithmic color scale. Shades of red denote less than 1\% error. The FT-IMSRG results display the most accuracy for higher $A$ and lower $N$.}
    \label{fig:cplotAN}
\end{figure}

We now turn to cases with larger values of $A$ and $N$, once again computing occupation numbers in the canonical ensemble.
In Fig. \ref{fig:cplotAN}, we show a the relative error in internal energy in parameter space at $\beta=5.0$ and various values for $A$ and $N$. We see a similar pattern to before, where weaker couplings generally lead to better agreement between the FT-IMSRG and exact energies. 

In the plots on the left of Fig. \ref{fig:cplotAN}, we show the case where the single-particle states are half-filled. Other than convergence issues for some parameters at $A=8$ and $N=16$, these plots are quite similar to each other. We see improvement in the FT-IMSRG's accuracy for nearly all parameters when $N$ is fixed and $A$ increases. When $N$ is increased for fixed $A$, however, the relative error increases slightly for most parameters. This can be understood by noting that the P3H interactions are analogous to unregulated delta function potentials, since the two-body matrix elements do not fall off in strength as the number of single-particle states $N$ is increased. Therefore, increasing $N$ effectively makes the interactions ``harder'', resulting in larger errors. Unsurprisingly, this effect is much more pronounced for stronger couplings in the Hamiltonian.
Note that in applications of the FT-IMSRG to nuclei, $N$ will be increased with a fixed $A$ to converge the result with respect to the basis size. In contrast, for the P3H model the $N\rightarrow\infty$ limit is not well-defined without renormalization.

\begin{figure}[t]
    \centering
    \includegraphics[width=0.45\textwidth]{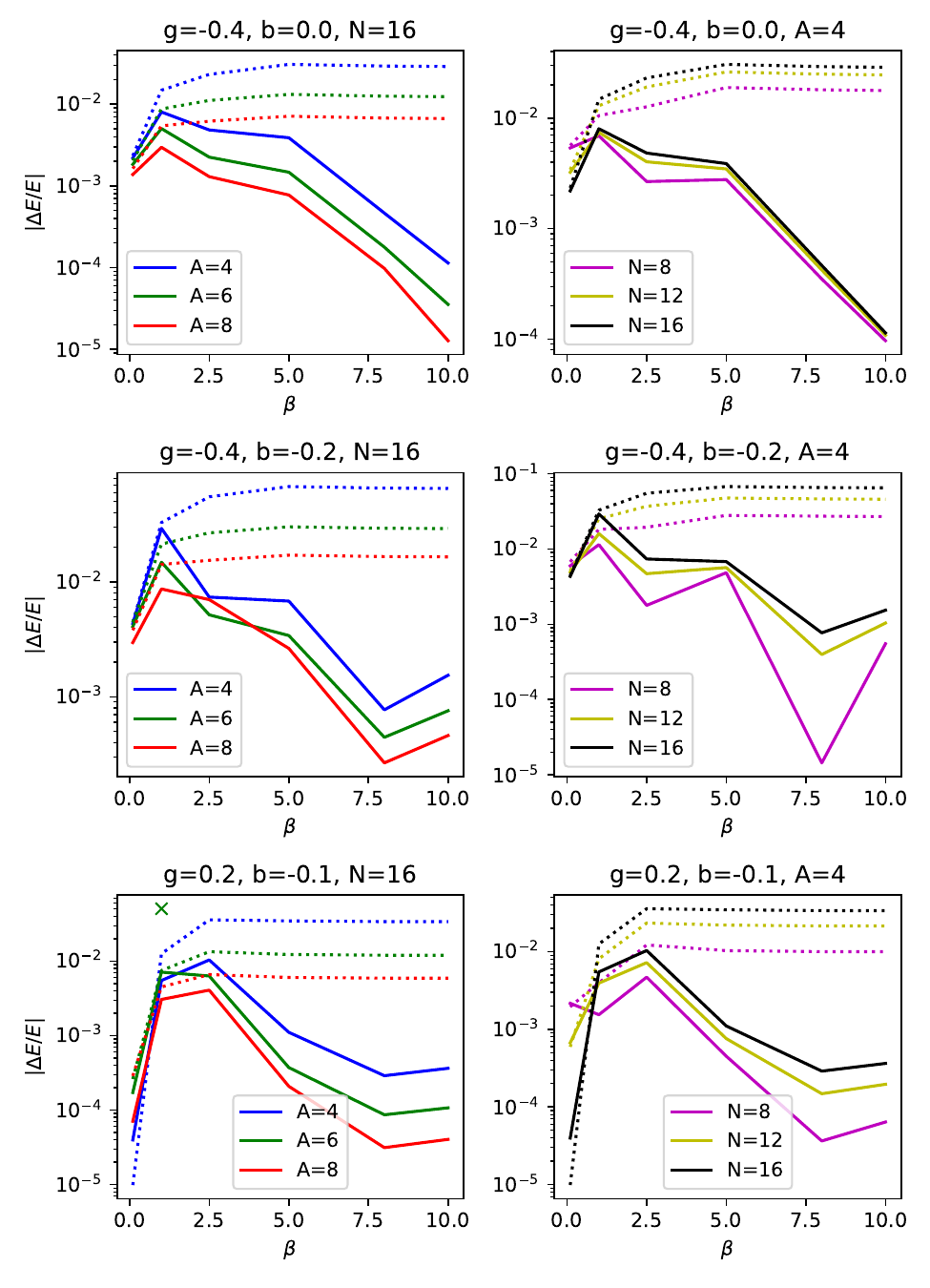}
    \caption{Relative error in internal energy versus inverse temperature on a logarithmic scale for various values of $g$, $b$, $A$, and $N$. The FT-HF results are shown with dotted lines and the FT-IMSRG results are shown with solid lines. In the bottom left panel ($g=0.2$, $b=-0.1$, $N=16$), we encountered numerical instability at $A=6$, $\beta=1$. The results for neighboring values of $\beta$ are plotted, with the unstable result at $\beta=1$ marked with an x. }
    \label{fig:logerrorAN}
\end{figure}

The behavior of the FT-IMSRG error with respect to changing $A$ and $N$ can be seen clearly in Fig. \ref{fig:logerrorAN}, which plots the relative error versus $\beta$ on a logarithmic scale for various values of $g$, $b$, $A$, and $N$. We see a significant increase in the FT-IMSRG's accuracy as $A$ is increased, and a smaller decrease in its accuracy as $N$ is increased. The FT-IMSRG typically improves on the FT-HF results, except for a few instances at high temperature/low $\beta$, which likely result from truncation errors. For the pure pairing interaction in the top row, the results are nearly identical as $N$ is increased -- this is because the pure pairing interaction couples fewer single-particle states than when the pair-breaking interaction is added  \cite{Davison:2023tx}. It is also notable in the bottom two rows that as $N$ is increased, the errors appear to converge from below to a fixed value. Since we would expect the error to converge from above as $N$ increases, we suspect that this behavior is caused by the use of the zero-range interaction. We will revisit this result in the future with a renormalized zero-range interaction.

\subsection{Comparing the Canonical and Grand Canonical Ensembles} \label{sec4d}

As mentioned previously, while it is a more accurate description of nuclei at finite temperature, the canonical ensemble is too computationally expensive to use for general realistic applications, but we can compute the canonical occupation numbers for the P3H model. Thus, we can explore the differences between canonical and grand canonical FT-HF and FT-IMSRG results for different model parameters to gain some insight for future applications.

\begin{figure}[t]
    \centering
    \includegraphics[width=0.45\textwidth]{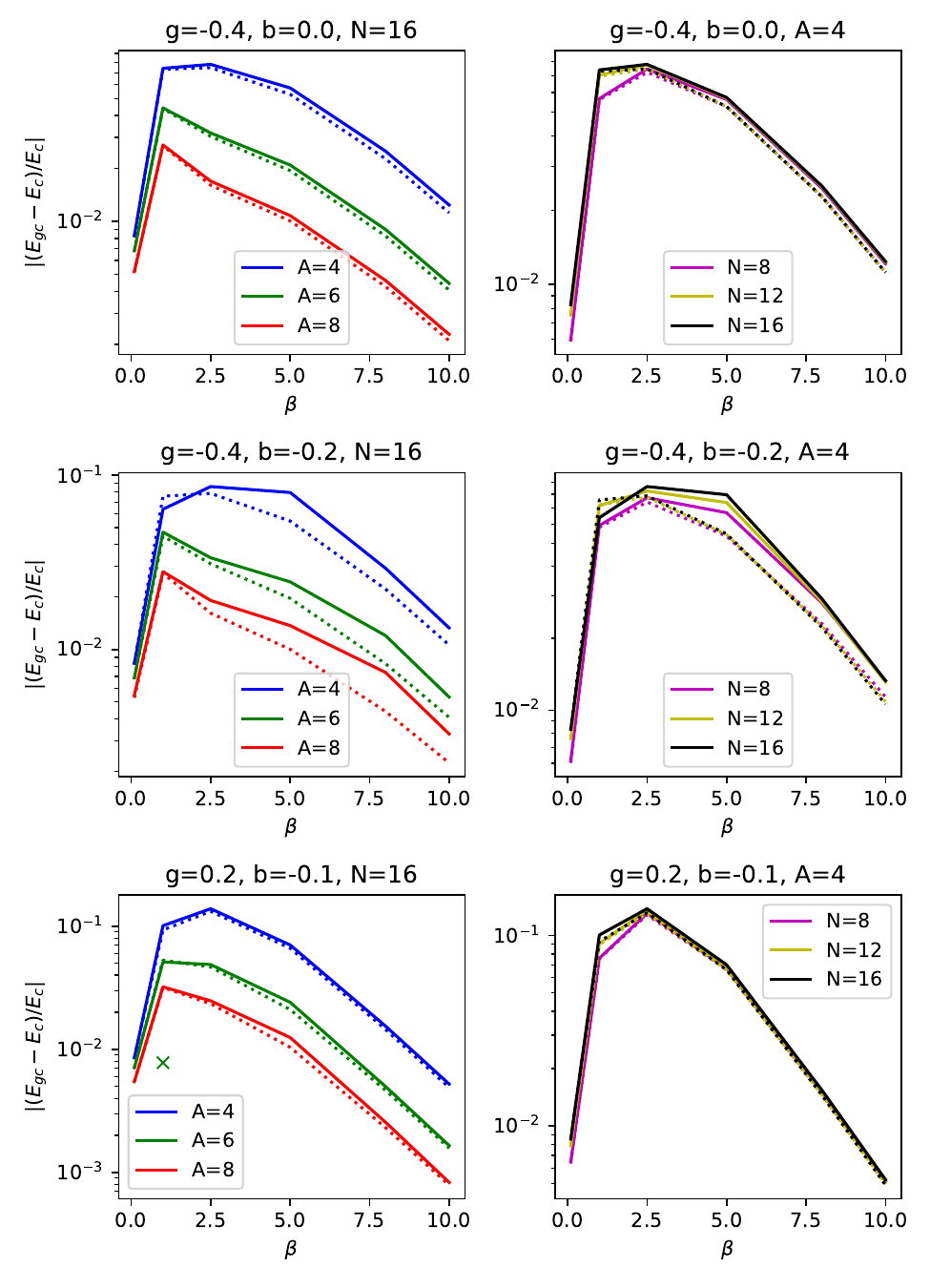}
    \caption{Relative difference between the canonical ensemble and grand canonical ensemble results versus inverse temperature on a logarithmic scale, using the same parameters as Fig. \ref{fig:logerrorAN}. The FT-HF results are shown with dotted lines and the FT-IMSRG results are shown with solid lines. As with Fig. \ref{fig:logerrorAN}, in the bottom left panel ($g=0.2$, $b=-0.1$, $N=16$) we encountered numerical instability at $A=6$, $\beta=1$ for the canonical ensemble. The results for neighboring values of $\beta$ are plotted, with the unstable result at $\beta=1$ marked with an x. }
    \label{fig:logerrorGCE}
\end{figure}

Figure \ref{fig:logerrorGCE} shows the relative difference between the canonical and grand canonical results for both FT-HF and FT-IMSRG. The differences are very similar in both methods, which is not surprising given that the choice of ensemble primarily affects the values of the occupation numbers prodcued by the FT-HF, which then serve as input for the FT-IMSRG, but remain unchanged during the flow. As a comparison of the center and bottom rows shows, the differences seem to be slightly more pronounced in FT-IMSRG than in FT-HF, which is likely the result of truncation effects.

As expected, there is a noticeable decrease in the difference between the ensembles upon increasing $A$ (see Sec. \ref{sec3} and \cite{Schonhammer:2017ms}). Changing $N$, however, has a significantly smaller effect on the comparison between the ensembles. This suggests that, as expected, the two ensembles will become equivalent in the thermodynamic limit, where $A$ and $N$ both become very large. For systems with $A\geq 8$, it seems safe to use the grand canonical ensemble, but for systems with fewer particles, this approximation does introduce an error in the 1-10\% range.

With all of this discussion, it is worth recalling that FT-IMSRG in the grand canonical ensemble does produce more accurate results than FT-HF in the grand canonical ensemble when compared to exact results. Thus, the FT-IMSRG remains successful as a post-HF method even in this approximation.

\subsection{Entropy and Free Energy} \label{sec4e}

Finally, we calculate entropy and free energy, quantities of much thermodynamic interest, from the FT-HF and FT-IMSRG results. Due to the FT-IMSRG evolution, the Hamiltonian has an implicit temperature dependence that is more complex than in the FT-HF case, and we cannot simply obtain $S$ by using Eqs.~\eqref{eq:gce_fthf_entropy} or \eqref{eq:ce_fthf_entropy}. Following the general idea of \cite{Harsha:2022dw}, we compute the entropy via integration. After performing the FT-IMSRG evolution for different values of $\beta$, we can express the internal energy as a function $E(\beta)$. This can be inverted to give $\beta$ as a function $\beta(E)$. We then have
\begin{equation}\label{entropy}
    S(\beta) = \int_{E(\infty)}^{E(\beta)} \beta'(E')dE'.
\end{equation}
Since the integrand $\beta(E)$ would be infinite in the zero-temperature limit, but the FT-IMSRG energy is insensitive to variations of $\beta$ in this regime, we use $E(\infty) \approx E(10)$, which gives an excellent (and controllable, if necessary) approximation.

\begin{figure}[t]
    \centering
    \includegraphics[width=0.45\textwidth]{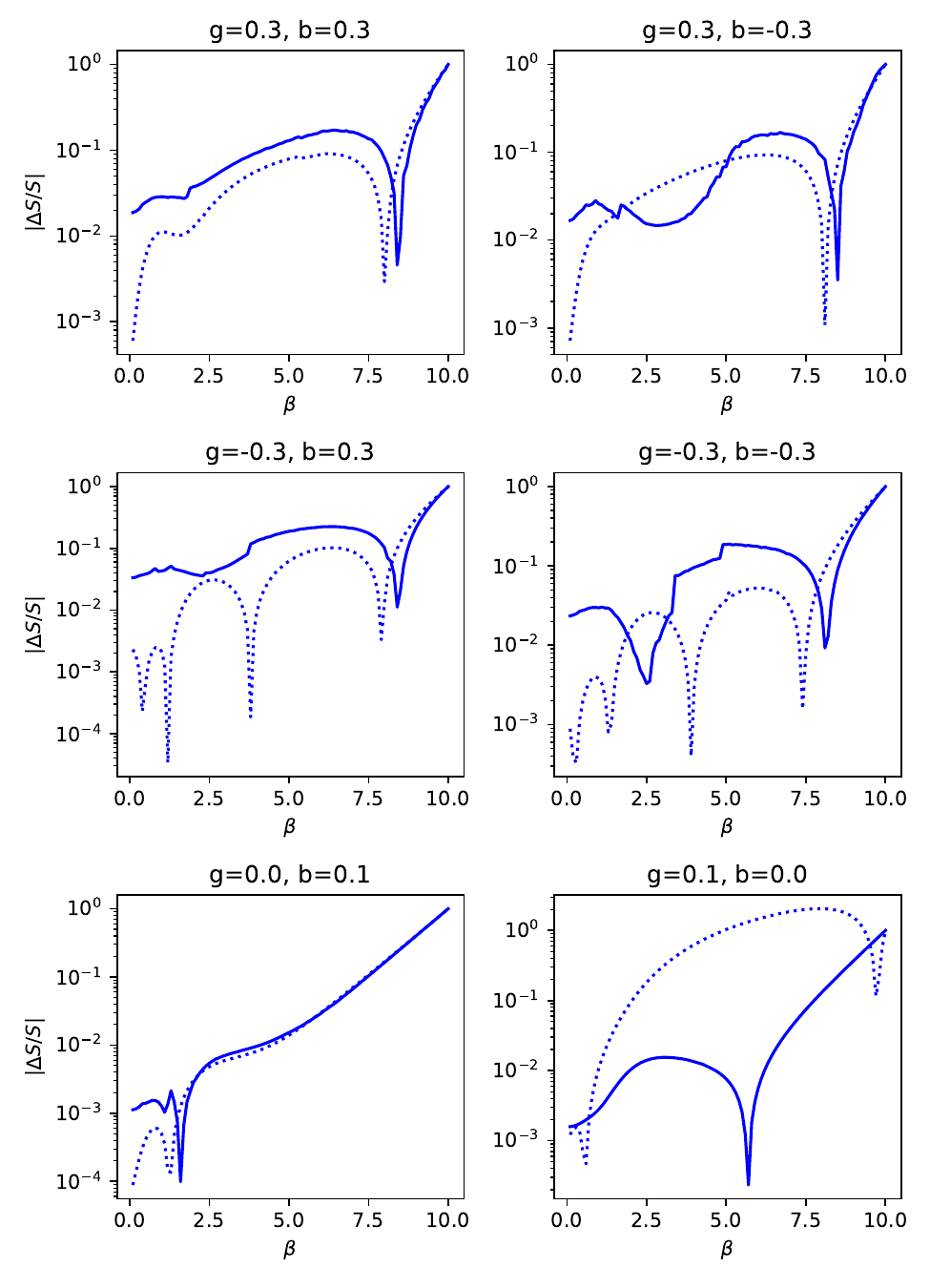}
    \caption{Relative error in entropy versus inverse temperature for $A=4$ and $N=8$ with various values of $g$ and $b$ on a logarithmic scale. The FT-HF results are shown with dotted lines and the FT-IMSRG results are shown with solid lines. The 100\% error at $\beta=10$ is a result of assuming $E(\infty)=E(10)$ in the integration limits of Eq. \eqref{entropy}. Entropy results are more accurate for weaker coupling, and the difference between FT-IMSRG and FT-HF in terms of entropy is generally small.}
    \label{fig:entropy}
\end{figure}

Fig. \ref{fig:entropy} shows the relative error in entropy for both FT-HF and FT-IMSRG, going back to the $A=4, N=8$ model. Notice that the $E(\infty)\approx E(\beta_\mathrm{max})$ scheme necessitates a 100\% error in the entropy at $\beta = \beta_\mathrm{max}$, which is seen in the figure. Outside of $\beta$ close to 10, the entropy calculations for both FT-HF and FT-IMSRG hover around 5-10\% error, except for the case of weak coupling. This is very similar to what was found for Finite-Temperature Coupled Cluster calculations~\cite{White:2020dn}.

\begin{figure}[t]
    \centering
    \includegraphics[width=0.45\textwidth]{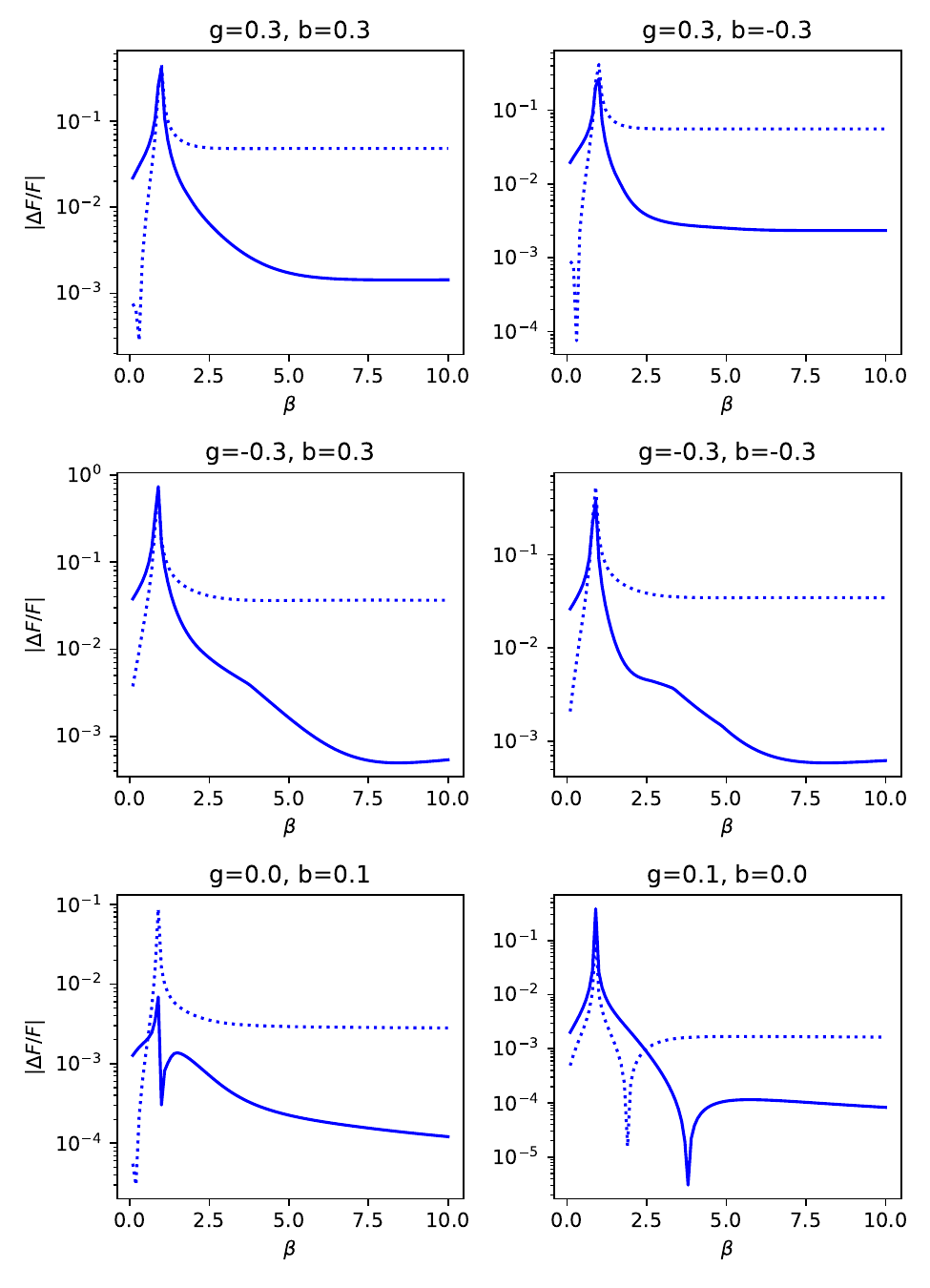}
    \caption{Relative error in Helmholtz free energy versus inverse temperature for $A=4$ and $N=8$ with various values of $g$ and $b$ on a logarithmic scale. The FT-HF results are shown with dotted lines and the FT-IMSRG results are shown with solid lines. Beyond $\beta\approx 2$, the FT-IMSRG results are consistently more accurate than the FT-HF results.}
    \label{fig:helmholtz}
\end{figure}

The Helmholtz free energy $F$ can then be calculated via Eq. \eqref{helm}. Fig. \ref{fig:helmholtz} shows the relative error in $F$ for both FT-HF and FT-IMSRG (once again with $A=4$ and $N=8$). We find that the FT-IMSRG calculation of $F$ significantly improves upon that of FT-HF, which is expected as this is heavily influenced by the accuracy of the internal energy calculations. Thus it is sensible that the relative error in $F$ resembles the relative error in $E$ (see Fig. \ref{fig:logerrorA4N8}).

The free energy is of significant importance because of its relationship to the canonical partition function $Z_A$ itself:
\begin{equation}
    \ln Z_A = -\beta F.
\end{equation}
This allows most thermodynamic properties of interest to be computed solely in terms of $F$, $\beta$, and derivatives of $F$. As we have shown that the FT-IMSRG can accurately calculate $F$, it can be used to reliably calculate these ensemble averages.

\section{Conclusion} \label{sec5}
In this work, we have extended the IMSRG to finite temperature, and demonstrated that the FT-IMSRG is a useful tool for calculating properties of many-fermion systems at finite temperature. Using the schematic model pairing-plus-particle-hole model that captures essential features of nuclear interactions, we performed a thorough assessment of the properties of FT-HF and FT-IMSRG, setting the stage for realistic applications of the FT-IMSRG to nuclei. We found the best agreement between FT-IMSRG and exact solutions at low temperatures and with weak coupling in the Hamiltonian, but the FT-IMSRG produced highly accurate results for a wide range of parameters and temperatures, improving on FT-HF.

As we looked at models with various different particle numbers $A$ and single-particle basis sizes $N$, we showed that the FT-IMSRG results improve in accuracy with higher $A$. We then demonstrated that the choice between the canonical and grand canonical ensembles in the setup of the FT-HF optimized occupation numbers can have noticeable effects on the FT-IMSRG results, but these effects become significantly lessened as the particle number increases. Finally, we used the FT-HF and FT-IMSRG results to calculate entropy and free energy, showing that the FT-IMSRG produces accurate results.

As our next steps, we will perform FT-IMSRG calculations for atomic nuclei with modern nuclear interaction derived from chiral Effective Field Theory. We will investigate the evolution of nuclear structure features like the neutron driplines with increasing temperature, and compute reaction and decay rates that are relevant for understanding nuclear processes in stellar environments, including nucleosynthesis. In parallel, we will pursue the implementation of finite temperature in other IMSRG variants and IMSRG-based hybrid methods \cite{Hergert:2017kx,Gebrerufael:2017fk,Stroberg:2019th,Yao:2020mw,Tichai:2023fl}.

\section*{Acknowledgements}
We thank A. Ravli\'{c} for useful discussions.

This work has been supported in part by the U.S. Department of Energy, Office of Science, Office of Nuclear Physics under Award Number DE-SC0023516, as well as the Jeffrey R. Cole Honors College Research Fund at Michigan State University. SKB is partially supported by National Science Foundation (NSF) Grants PHY-2013047 and PHY-2310020.

\medskip

\bibliographystyle{apsrev4-1}
\bibliography{refs}

\end{document}